\documentclass[11pt,reqno]{amsart}

\usepackage{amsmath, amsfonts, amssymb, amscd, enumerate, mathrsfs}
\usepackage{graphicx}
\usepackage{cyrillic}

\newcommand{\cyrit}{\fontencoding{OT2}\selectfont\textcyrit}

\newcommand{\cyrbf}{\fontencoding{OT2}\selectfont\textcyrbf}

\theoremstyle{plain}
\newtheorem{theo}{Theorem}[section]
\newtheorem{lem}[theo]{Lemma}

\theoremstyle{definition}

\newtheorem{example}[theo]{Example}
\newtheorem{definition}[theo]{Definition}

\theoremstyle{plain}

\theoremstyle{definition}

\renewcommand{\=}{\overset{\operatorname{def}}{=}}

\newcommand{\beq}{\begin{equation}}
\newcommand{\eeq}{\end{equation}}
\renewcommand{\a}{\alpha}
\renewcommand{\b}{\beta}

\renewcommand{\d}{\delta}
\newcommand{\e}{\epsilon}

\newcommand{\h}{\eta}

\renewcommand{\l}{\lambda}
\newcommand{\m}{\mu}
\renewcommand{\o}{\omega}

\newcommand{\s}{\sigma}

\newcommand{\D}{\Delta}

\newcommand{\G}{\Gamma}

\renewcommand{\L}{\Lambda}

\newcommand{\bA}{\mathbb{A}}

\newcommand{\bC}{\mathbb{C}}
\newcommand{\bD}{\mathbb{D}}
\newcommand{\bF}{\mathbb{F}}
\newcommand{\bR}{\mathbb{R}}
\newcommand{\bZ}{\mathbb{Z}}

\newcommand{\bH}{\mathbb{H}}
\newcommand{\bK}{\mathbb{K}}
\newcommand{\bN}{\mathbb{N}}

\newcommand{\ga}{\mathfrak{a}}
\newcommand{\gb}{\mathfrak{b}}

\newcommand{\gf}{\mathfrak{f}}
\renewcommand{\gg}{\mathfrak{g}}
\newcommand{\gh}{\mathfrak{h}}

\newcommand{\gp}{\mathfrak{p}}

\newcommand{\gF}{\mathfrak{F}}
\newcommand{\gX}{\mathfrak{X}}
\newcommand{\gJ}{\mathfrak{J}}

\newcommand{\so}{\mathfrak{so}}

\newcommand{\ggl}{\mathfrak{gl}}

\newcommand\SO{\mathrm{SO}}

\newcommand\Spin{\mathrm{Spin}}

\newcommand{\cA}{\mathcal{A}}
\newcommand{\cB}{\mathcal{B}}
\newcommand{\cC}{\mathcal{C}}
\newcommand{\cD}{\mathcal{D}}
\newcommand{\cE}{\mathcal{E}}
\newcommand{\cF}{\mathcal{F}}
\newcommand{\cG}{\mathcal{G}}
\newcommand{\cH}{\mathcal{H}}

\newcommand{\cL}{\mathcal{L}}

\newcommand{\cR}{\mathcal{R}}
\newcommand{\cS}{\mathcal{S}}
\newcommand{\cT}{\mathcal{T}}
\newcommand{\cU}{\mathcal{U}}
\newcommand{\cV}{\mathcal{V}}

\newcommand{\cZ}{\mathcal{Z}}

\newcommand\Dirac[3]{\Gamma_{#1#2}^{\ #3}}

\renewcommand{\square}{\kern1pt\vbox
{\hrule height 0.6pt\hbox{\vrule width 0.6pt\hskip 3pt
\vbox{\vskip 6pt}\hskip 3pt\vrule width 0.6pt}\hrule height0.6pt}\kern1pt}

\DeclareMathOperator\Aut{Aut\;}
\DeclareMathOperator\aut{aut}
\DeclareMathOperator\Ad{Ad}
\DeclareMathOperator\ad{ad}

\DeclareMathOperator\Id{Id}

\DeclareMathOperator{\Der}{Der\;}

\DeclareMathOperator{\Span}{Span}

\renewcommand\Im{\operatorname{Im}}

\newcommand{\Hom}{{\operatorname{Hom}}}
\newcommand{\Sym}{{\operatorname{\it{Sym}}}}

\newcommand{\wt}{\widetilde}
\newcommand{\wh}{\widehat}

\newcommand{\sheaf}{{\operatorname{\it Sheaf}}}

\newcommand{\n}{\nabla}

\newcommand{\dl}{|\!|}

\newcommand{\LL}{{{\cyrit{L}}\!\!\operatorname{\it l}}}
\newcommand{\LLbis}{{\cyrbf{L}}}

\newcommand{\be}{\begin{equation}}
\newcommand{\ee}{\end{equation}}

\def\<#1,#2>{\langle\,#1,\,#2\,\rangle}
\newcommand{\arr}{\begin{array}{rlll}}
\newcommand{\ea}{\end{array}}
\newcommand{\bea}{\begin{eqnarray}}
\newcommand{\eea}{\end{eqnarray}}
\newcommand{\bean}{\begin{eqnarray*}}
\newcommand{\eean}{\end{eqnarray*}}

\def\sideremark#1{\ifvmode\leavevmode\fi\vadjust{
\vbox to0pt{\hbox to 0pt{\hskip\hsize\hskip1em
\vbox{\hsize3cm\tiny\raggedright\pretolerance10000
\noindent #1\hfill}\hss}\vbox to8pt{\vfil}\vss}}}

\newcounter{ssig}
\newcounter{ttig}
\newcommand\sect[1]{\G({#1})}
\newcommand\locsect[1]{\G_{\operatorname{loc}}({#1})}

\newcommand\Ricperp{\operatorname{Ric}^{\cD^\perp}}

\title[Super-Poincar\`e algebras, space-times  and  supergravities (II)]
{Super-Poincar\`e algebras, \\
space-times and  supergravities (II)}

\author{A. Santi and A. Spiro}

\address{
Andrea Santi,
Facult\'e des Sciences, de la Technologie et de la Communication,
Universit\'e du Luxembourg,
L-1359 Grand-Duchy of Luxembourg.}
\email{andrea.santi@uni.lu}
\address{
Andrea Spiro, Scuola di Scienze e Tecnologie, Universit\`a di Camerino, 
Camerino, 
Italy.}
\email{andrea.spiro@unicam.it}
\thanks{\\ \phantom{ccc}The first author was supported by project F1R-MTH-PUL-08HALO-HALOS08 of University of Luxembourg.}

\keywords{Supermanifolds; Eleven Dimensional Supergravity; Principle of General Covariance; Super-Poincar\`e algebras}
\subjclass[2000]{83E50, 58A50, 17B70.}

\begin{document}

\begin{abstract} The presentation of supergravity theories of our previous paper  ``Super-Poincar\`e algebras, space-times  and  supergravities (I)''  is  re-formulated in  the language  of Berezin-Leites-Kostant theory of supermanifolds.  It is  also shown that the equations of Cremmer, Julia and Scherk's theory of 11D-supergravity are equivalent to   manifestly covariant equations on a supermanifold. 
 \end{abstract}
 
\maketitle

\null \vspace*{-.25in}

\section{Introduction}
\setcounter{section}{1}
\setcounter{equation}{0}
In a previous article (\cite{SaS}), we proposed   formulations of  supergravity theories, 
based on the notion of an  extended space-time  $(M, M_o, \cD)$,  formed by a superspace $M$,  a distinguished 
submanifold $M_o \subset M$ (representing the physical  space-time)  and  a non-integrable distribution $\cD$, with properties determined by imposed supersymmetries of   vacuum solutions.  In such formulations, a supergravity theory  is  represented by 
a  collection of tensor fields and tensorial  equations  on $M$,  whose restrictions at the points of $M_o$ give  the  physical   fields  and their equation of motion.  \par
\medskip
Many  conceptual ingredients of our  approach, such as the notion of  a superspace  $M$ and   
of  tensorial equations on a superspace,  are  standard. 
But, at the best  of our knowledge,  a presentation  of  all   material  in a coordinate-free language and in terms of classical   differential geometric objects  was still  missing.
So, we developed such presentation  pursuing the following leading intent: {\it Get an  economical description 
of  supergravity theories in terms of  objects,  which can  be studied  with   standard   techniques of Differential Geometry}. \par
\medskip
Let us  briefly recall the main points  of our  presentation    in \cite{SaS}. \par
Given a Poincar\`e algebra $\gp = \so(V) + V$ of a flat pseudo-Riemannian space $V = \bR^{p,q}$ and a (super or $\bZ_2$-graded) extended algebra $\gg = \so(V) + V + S$,  we call  {\it space-time of type $\gg$} any (super-)manifold $M$,  together   with a  distinguished submanifold $M_o \subset M$ and  a non integrable distribution $\cD$, whose    Levi form $\cL$ is modeled on the Lie brackets of elements in  $S \subset \gg$.  \par
 A 
 {\it gravity field on $M$} is a pair $(g, \n)$,  formed by a   tensor field $g$  of type $(0,2)$,   inducing a pseudo-Riemannian metric  on the  $g$-orthogonal distribution $\cD^\perp$,  and  a covariant derivation    $\n$ preserving   $\cD$, $g$ and $\cL$. \par
 A  {\it supergravity of type $\gg$} is  a pair $\cG = ((M, M_o, \cD), (g, \n))$,  formed by a space-time  $(M, M_o, \cD)$ of type $\gg$ and a gravity field $(g, \n)$.  The  {\it physical fields} of $\cG$ are  formed by a pair of  covariant derivations on $M_o$,  called  {\it metric} and {\it spinor connections}, and by  three tensor fields, representing  the {\it graviton}, the {\it gravitino} and the {\it auxiliary field(s)}, respectively. \par
 \smallskip
There is a distinguished class of supergravities,  the so-called  {\it (strict) Levi-Civita supergravities of type $\gg$}, characterized  by the vanishing  of  some special parts of the  torsion $T$ of   $\n$. The importance of such supergravities comes  from an   existence and uniqueness theorem,   which implies that  their physical fields 
are completely determined by the graviton, the gravitino and the auxiliary field(s), as it is  required   in the standard component approach to supergravities.  \par
We also recall that in \cite{SaS},  we  showed that the  variations of   graviton and gravitino  of a Levi-Civita supergravity,  determined  by  Lie derivatives along vector fields of  $M$,  nicely match the supersymmetric transformation  rules of simple 4D-supergravity and other  supergravities, determined  in component formalism. And one can directly check that the same  occurs for the variations of graviton and gravitino in Cremmer-Julia-Scherk 11D-supergravity (\S \ref{section52}). All this  can be considered as a supporting evidence for   the idea that    (localized) supersymmetric invariance of supergravity theories  is actually  a sort  of  {\it  Principle  of  General Covariance}, i.e.    a principle of  invariance under local changes of coordinates (or, equivalently, local diffeomorphisms) of the superspace  $M$. \par
\smallskip
The purpose of this  paper is to rewrite the contents of  \cite{SaS}  in terms of  well defined notions of supergeometry.  
We   essentially follow  the   theory  developed by  Berezin, Leites, Kostant et al., but with a modified  notion  of   supermanifold, which  we call  $\LL$-supermanifold (\S \ref{section2.2}; see also  \cite{Sc}).  
A  $\LL$-supermanifold $M^\LL = (M_o, \cA_M^\LL)$  is characterized by  an algebra  $\cA_M^\LL$ of superfunctions,  which can be considered as  the  super-analogue of an algebra  on a smooth manifold of smooth functions taking values into  a  $\bZ_2$-graded algebra (and hence generated by  ``even''  and ``odd-valued''  functions).  In fact,  the need for  $\LL$-valued functions naturally arises in any theory
involving  fermions (anti-commuting quantum fields), not only 
supersymmetric ones.  In  supergravity theories bosons and 
fermions are necessarily intertwined; there is therefore even more reason 
to consider $\LL$-valued superfunctions and $\LL$-supermanifolds.
\par
\smallskip
After this, we apply our approach   to the case  of Cremmer, Julia and Scherk's theory of  $11D$-supergravity  (\cite{CJS}). 
More precisely, we show how the fields and equations of $11D$-supergravity can be   expressed    in terms of a supergravity $\cG = ((M^\LL, M^\LL_o, \cD), (g, \n))$ of type $\gg$ endowed with a suitable 4-form $\cF$ on $M^\LL$.  Using  results of \cite{CDF}, we  get the existence of  a  one-to-one correspondence between 
  solutions in component formalism   of $11D$-supergravity equations  and    quadruples $(\cD, g, \n, \cF)$ on $M^\LL$, 
  which satisfy a  set of constraints and equations of purely tensorial type. Due to  this, all questions  concerning  constructions and analysis  of  solutions of  $11D$-supergravity can be  naturally reduced   to  problems    on suitable geometric structures   on $\LL$-supermanifolds. \par 
\medskip
The structure of the paper is the following. In \S 2, we use  tensor products  with a suitable exterior algebra $\LL = \L^* W$ to define objects that behave as   even/odd valued  functions and even/odd vector  fields   on a classical smooth manifold. In \S 3,  we introduce  the definition of ``$\LL$-supermanifold'', which is   a supermanifold with an algebra of superfunctions, analogous  to  the even/odd valued functions on a  manifold. First properties of $\LL$-supermanifolds are given;  In particular,  we show that the  Lie derivatives  along (even) super vector fields are related with 1-parameter groups of (local) diffeomorphisms  in perfect analogy with the corresponding  relation between Lie derivatives and flows  of  smooth manifolds. In \S 3, we re-formulate  definitions and  properties  of supergravities of type $\gg$   in the language of $\LL$-supermanifolds. In \S 4, we show how Cremmer, Julia and Scherk's theory of eleven dimensional supergravity can be encoded  as a theory on supergravities of type $\gg$ and that it satisfies the {\it generalized Principle of General Covariance}, introduced  in  \cite{SaS}. 
For reader's convenience,  we  briefly outline   the theory of supermanifolds   in  the Appendix.   \par
\medskip
We conclude observing that  the generalization of the notion of supermanifolds, considered in this paper,  stems naturally    from  a functorial approach to    supergeometry as considered for instance in \cite{Mo, He1,Sa, SW} (see also \cite{Fr}, Lec. 2). We believe  that  this is indeed   the most  appropriate approach to supergeometry.   
\par
\medskip
\subsubsection*{Notation}
Given a sheaf or bundle $\pi: \cA \to N$ over a manifold $N$, we denote by  $\cA_{\dl\cU}$ the restriction of  $\cA$   over a subset $\cU \subset N$. In particular  $\cA_{\dl x} = \pi^{-1}(x)$ for any $x \in N$.  The set   of  global (resp. local)  smooth sections of $\cA$  is  denoted by $\sect \cA$   (resp.  $\locsect \cA$).  The  sheaf of germs of  sections of a bundle $\cA$  is denoted by   $\sheaf \cA$. 
Given a manifold $N$,  we denote by $\gX(N) = \sect{TN}$  the class of smooth vector fields 
and by $\gF_{N}$ the sheaf of germs of smooth real functions of $N$.  \par
\smallskip
If  $X$ is  a  derivation of a ring $R$, its action  on  elements $f \in R$ is denoted   either by $X (f)$ or  by $X \cdot f$.  If  $\cA = \cA_0 + \cA_1$ is a   $\bZ_2$-graded vector space, the  parity $i  = 0, 1$ of an  homogeneous element $\gf \in \cA_i$, is  denoted by $|\gf|\in\bZ_2$.\par
\smallskip
We consider  Clifford algebras as defined   e.g. in \cite{LM} and  the Clifford product   between  vectors   of the  standard basis of $\bR^{p,q}$ is  $e_i \cdot e_j = -  2 \eta_{ij}$ and  not  `` $+  2 \h_{ij}$'', as it is  often assumed  in  Physics. Due to this,   our notation for  signatures of Clifford algebras  is  opposite to the one of several  other   papers.   \par
\bigskip
\section{$\LLbis$-valued functions  on  manifolds and $\LLbis$-supermanifolds}
\setcounter{equation}{0}
\label{section2}
\subsection{$\LLbis$-valued functions and $\LLbis$-valued  fields  on  classical manifolds}\label{2.1}
\hfill\par
In Quantum Field Theory,   fields of  bosonic or  fermionic particles  correspond to  commuting  or 
anti-commuting operators on some suitable Hilbert space. This fact forces to represent them as 
 differential geometric objects,  constructed  with 
the following defined {\it graded}  functions and tensors. \par
\smallskip
Given a finite dimensional  vector space $W = \bK^N$, $\bK = \bR$ or $ \bC$,  we denote by  $\LL \= \bigoplus_{i = 0}^N \L^i W$ its  exterior algebra, with  spaces of {\it even} and {\it odd} elements given by  
$$ \LL_0 \= \bigoplus_{i = 2 p} \L^i W \ ,\qquad \LL_1 \=  \bigoplus_{i = 2 p + 1} \L^i W\ .$$
Elements $\h \in \LL_i$, $i = 0,1$,  are called {\it homogeneous of parity $|\h| = i$}. \par
\smallskip
The consideration of an exterior algebra $\LL = \bigoplus_{i = 0}^N \L^i W$ allows constructions  of appropriate  models for  quantum  anti-commuting   fields, {\it provided that $\dim W$ is large enough to prevent unwanted cancellations in products}.   In this paper, this is actually the only condition we have to worry about and,  from now on,   {\it we   consider $W = \bK^N$ as  fixed,    with $N$  sufficiently large}. \par
\medskip
Let $M_o$ be   a smooth manifold   and $E = E_0 + E_1  \longrightarrow M_o$ a real (resp. complex) $\bZ_2$-graded vector bundle of finite rank, with    fiber $V = V_0 + V_1$. The   subbundles $\pi_i:E_i \longrightarrow M_o$, $i = 0,1$, are  called  {\it subbundles of even/odd elements}, respectively.
 We remark   that     bundles,  in which either $E_1$ or $E_0$  is  trivial,   are admissible: We call them  {\it (purely) even} or {\it (purely) odd bundles}, respectively. Clearly,  any  ungraded vector bundle $\pi: F \longrightarrow M_o$ might be endowed with parity equal to $0$ or $1$ and hence considered as an even or odd,  according to the needs. \par
\bigskip
Given  $M_o$ and a $\bZ_2$-graded vector bundle $\pi: E = E_0 + E_1 \longrightarrow M_o$,  we denote by $M^{\LL}_o$  and $E^{\LL}$ the   bundles  
$$\pi_o^\LL: M^{\LL}_o \= \LL \times  M_o\longrightarrow M_o\ ,\qquad \pi^\LL: E^{\LL} \= \LL \otimes_\bK E \longrightarrow M_o\ ,$$
with   $\LL = \L^* W$ (here   $W = \bR^N$  if    $E$ is  real and $W = \bC^N$  if  $E$ is complex) and by
$\LL \otimes_\bK E$  the bundle with fibers $\LL \otimes_\bK E_{\dl x}$.
Sections of  $\gF^{\LL}_{M_o} \= \sheaf{(M_o^{\LL})}$ are called  {\it  $\LL$-valued functions}, while    $E^{\LL}$ is called the {\it lambdification of $E$}.\par
\smallskip
The bundles $M^\LL_o$ and $E^\LL$ have natural structures of $\bZ_2$-graded bundles 
$$M_o^\LL = M_o^{\LL_0} + M_o^{\LL_1}\ ,\qquad E^\LL = (E^\LL)_0 +  (E^\LL)_1\ ,$$ 
where   $M_o^{\LL_i} = \LL_i \times M_o$, 
$(E^\LL)_0 =\LL_0 \otimes E_0 + \LL_1 \otimes E_1$ and $(E^\LL)_1 = \LL_1 \otimes E_0 + \LL_0 \otimes E_1$.
The (local) sections of $M_o^{\LL_0}$ (resp. $M_o^{\LL_1}$) are called {\it even} (resp. {\it odd}) {\it valued functions}, while 
the (local) sections of $(E^\LL)_0$ (resp. $(E^\LL)_1$) are called  {\it even} (resp. {\it odd}) {\it sections}.\par
\smallskip
Given a  $\LL$-function $f=\sum_\a \eta_{\a}\otimes f_\a\in\locsect{M_o^\LL}$ and a section $X=\sum_\a \eta_\a\otimes X_\a\in\locsect{E^{\LL}}$,  we call {\it evaluations at $x\in M_o$}  the values
$$
f|_{x}\=\sum_\a \eta_{\a}f_\a(x)\in\LL\quad , \quad X|_{x}\=\sum_\a \eta_\a\otimes X_\a(x)\in E_{\dl x}^\LL = \LL\otimes_\bK E_{\dl x}\ .
$$
$\sheaf{(E^{\LL})}$ has a natural structure of locally free sheaf of $\gF^{\LL}_{M_o}$-moduli, with  products  between sections of $\gF^{\LL}_{M_o}$ and $ \sheaf{(E^{\LL})}$   defined by
$$  (\h \otimes f) \cdot (\h'\otimes e) \= \h\h' \otimes(f \cdot e)\ ,$$
for any $\h, \h' \in \LL$, $ f \in \sect{\gF_{M_o}}$, $e \in \sect{E}$.
\par
\medskip
Given a smooth map $\varphi: M_o \longrightarrow N_o$, the pull-back $\varphi^*: \gF_{N_o} \longrightarrow \gF_{M_o}$ induces a corresponding   morphism of sheaves of $\LL$-moduli $\varphi^*: \gF^{\LL}_{N_o}\longrightarrow \gF^{\LL}_{M_o}$. Similarly, any even bundle morphism $\Phi: E \longrightarrow E'$, 
induces an even $\LL$-linear bundle morphism 
$\Phi: E^{\LL} \longrightarrow E'{}^{\LL}$.
\par
\begin{definition} Given an open subset $\cU_o \subset M_o$, we call {\it  frame field of $E^\LL$ on $\cU_o$}  any  collection $(e_1, \dots, e_r)$ of homogeneous $e_i \in \sect{E^\LL_{\dl \cU_o}}$ such that: 
\begin{itemize}
\item[--] $(e_1|_x, \dots, e_r|_x)$ is a basis of the $\LL-$module $E_{\dl x}^\LL$ for any $x \in \cU_o$; 
\item[--]  there is an integer  $r_o \in \bN \cup \{0\}$, such that the $e_i$'s, with   $1 \leq i \leq r_o$,  are in  $\sect{(E^\LL_{\dl \cU_o})_0}$, while the $e_j$'s, with  $r_o+1\leq j \leq r$,  are in  $\sect{(E^\LL_{\dl \cU_o})_1}$. 
\end{itemize}
The elements $(e_1, \dots, e_{r_o})$ (resp. $(e_{r_o+1}, \dots, e_{r})$) are called {\it even} (resp. {\it odd}) {\it elements of the frame field}. \par
\end{definition}
The number of even and odd elements of a local frame field does not depend on the choice of the local frame, due to the following simple lemma, whose proof is left to the reader. \par 
\begin{lem} For any $x \in M_o$,    $ \dim_{\LL} E^\LL_{\dl x} =   \dim_\bK E_{\dl x}$ and  the number of  even elements  of a frame field $(e_i)$  is    $r_o =  \dim_\bK E_0{}_{\dl x}$.
\end{lem} 
Local frames allow the definition of  the following subbundles.  Given $0 \leq p_o \leq r_o$, $0 \leq q_o \leq r - r_o$,   a subbundle $F \subset E^\LL$  is called {\it regular of bi-rank $(p_o, q_o)$} if any $x_o \in M_o$ admits a neighborhood $\cU_o$ and a  frame field $(e_i)$ on $\cU_o$, such that 
$$F_{\dl x"}  = \Span_\LL \left(e_j|_{x}, \  j \in \{ 1, \dots, p_o\} \cup\{r_o + 1, \dots, r_o + q_o\}\ \right)\ ,\qquad \  x \in \cU_o\ .$$\par 
\medskip
Coming back to tuples $(\psi^\a)$ of anti-commuting operators, which   locally   represent fermions  (as e.g. the components of  a Dirac  field), we may conveniently identify  them  with  odd-valued components of   sections  $\psi: \cU \longrightarrow E^\LL$ of a suitable lambdification  $E^\LL$. 
The parity  of   $\psi$  depends  on the parity of the  elements of   the  frame field. 
 In case $E$ is   decreed  to be purely  even (resp. odd), the frame fields has only even (resp. odd) elements and   $\psi$ has clearly odd-valued components $\psi^\a$ if and only if it is odd  (resp. even).
Both alternatives can be handled in   equivalent ways, but  we found the second  one easier-to-use. So,  from now on, we adopt  the following conventions. 
\par
\begin{itemize}
\item[--] A lambdification $E^\LL$ is  called  {\it bosonic} (resp.  {\it fermionic}) if   the underlying vector bundle $E$ is purely even (resp. odd); A  subbundle $F \subset E^\LL$ of a lambdification $E^\LL$ is called   {\it bosonic} (resp.  {\it fermionic})
if it is regular of bi-rank $(p_o, 0)$ (resp. $(0, q_o)$); 
\item[--] We  call  {\it  fields in $E^\LL$}  the {\it even} sections $\psi: \cU \longrightarrow (E^{\LL})_0$. A field  is  called  {\it bosonic} (resp.  {\it fermionic}) if  and only if  it is  an even section of  a bosonic (resp.  fermionic)  bundle. 
\end{itemize}
\par
We   conclude  with  the notion of ``conjugation''. Following a common  habit in Physics (\cite{CDF}, p. 336),  we call  {\it standard conjugation of $\LL = \L^* W$,  $W = \bC^N$},   the  unique $\bC$-anti-linear involution,  which coincides  on $W$ with the usual  conjugation and is  an algebra  anti-homomorphism, i.e. such that 
$\overline{w_1 \cdot w_2} = \overline{w_2}\cdot \overline{w_1}$ for any $w_i \in \LL$.  We use  the symbol ``\ $\overline{\phantom{a}}$\ ''  to denote also   the    associated standard conjugation of    (local)  sections of $M_o^\LL = \LL \times M_o$. \par
If 
 $\pi: E\longrightarrow M_o$ is a complex vector bundle, endowed   with a  $\bC$-antilinear, fiber bundle involution  
$\overline{\phantom{a}}: E \longrightarrow E$, we  use the same symbol  ``\ $\overline{\phantom{a}}$\ ''   to denote     the induced  involution on $E^\LL$, defined by 
 $$\label{conjugation}Ê \overline{ f \cdot e} = (-1)^{|f| |e|} \overline{f} \cdot \overline{e} \ \   \text{for any homogeneous} \ f  \in \locsect{\gF_{M_o}^\LL}, e \in \locsect{E}.$$

\begin{example}  We want to show how Dirac's    Lagrangian  for free electrons 
 (see e.g. \cite{BD}) can be defined using the objects of  this section.  Let $\pi: \cS = S \times \bR^{3,1}\longrightarrow \bR^{3,1}$ be the (trivial) spinor bundle of 
$ \bR^{3,1}$,  with fiber given by the space of Dirac spinors $S = \bC^4$.  {\it We consider $\cS$ as a  purely odd bundle}, so that   electrons (which are fermions!)  are  represented by  fields (=  even  sections) $\psi: \cU \subset \bR^{3,1}Ê \longrightarrow \cS^\LL$ in the fermionic bundle $S^\LL$.      Dirac's    Lagrangian can be considered as  the map
$\cL: \locsect{\cS^\LL} \longrightarrow \locsect{(\L^4 T^*\bR^{3,1})^\LL}$ given by 
$$\cL(\psi) = \left\{i \left< \overline \psi,    \G^j \cdot D_{e_j} \psi\right> - m\left<\overline{\psi},\psi\right> \right\}dx^0 \wedge dx^1 \wedge dx^2\wedge dx^3\ ,$$
where: 
\begin{itemize}
\item[--] $D$ is the flat Levi-Civita connection of $\bR^{3,1}$ and $(e_i)$ is the standard orthonormal basis of $\bR^{3,1}$,  with $<e_i, e_j> = \h_{ij}Ê= \e_i \d_{ij}$ where  $\e_0 = -1$ and $\e_1 = \e_2 = \e_3 = 1$; 
\item[--]  $\G^j \cdot : \cS^\LL \longrightarrow \cS^\LL$ are  the $\LL$-linear bundle maps,  represented in  standard  frames by the classical Dirac matrices
$\G^0 =     \left( \begin{matrix} I & 0 \\
0 & - I \end{matrix}\right)$ and $\G^j  = \left( \begin{matrix} 0  & \s^j \\
 -\s^j & 0  \end{matrix}\right) \  \text{for}\ j \neq 0$, 
with  $\s^1 =  \left( \smallmatrix 0 & 1 \\ 1 & 0\endsmallmatrix\right)$, $ \s^2 = \left(
\smallmatrix 0 & -i \\ i & 0\endsmallmatrix \right)$, $\s^3 = \left(
\smallmatrix 1 & 0 \\ 0  & -1\endsmallmatrix \right)$;
\item[--] $\left<\cdot, \cdot\right>: \cS^\LL \times \cS^\LL \longrightarrow  \LL$ is  the   $\LL$-bilinear  map along   fibers (symmetric on even sections),  defined by $\left< s_1,  s_2\right>|_x  =  s^T_1|_x  \cdot \G^0 \cdot s_2|_x$ and 
$m$ is the inertial mass of the electron. 
\end{itemize}
\end{example}
\medskip
\subsection{$\LLbis$-supermanifolds}\hfill\par
\label{section2.2}
\smallskip
 We recall that there exist
 two distinct approaches to  the notion of ``supermanifold'',  one developed by Berezin, Bernstein, Leites, Kostant and others (\cite{Be, BK, BL, Le, Ko, Ma}) and another invented by   DeWitt, Batchelor and Rogers
 (\cite{DeW, Ba, Ba1, Ro, Ro1, BBHR, Ro2}). These   approaches turn out to be  equivalent if  some technical   modifications and adjustments of  basic definitions are considered  (see  \cite{Ba1, Ro1, Sa}). \par
We   follow  Berezin-Leites-Kostant approach, of which the   reader can find all main  definitions and properties  in  the appendix.  But we have to stress the fact that,  in order to reach  a satisfactory and rigorous treatment of bosons and fermions,   one has to   consider supermanifolds with   analogues  of the even/odd valued functions defined  in  \S \ref{2.1}.  This  forces   to adopt the following modification of   Berezin-Leites-Kostant  definition of supermanifold (see also \cite{Sc, He1}). Notice that:
 \begin{itemize}
 \item[--]   such  modification    essentially  coincides with  the  adjustment that   makes Berezin-Leites-Kostant approach   equivalent to the Batchelor-DeWitt-Rogers approach and 
 \item[--]   it is  motivated also by Molotkov's categorical approach (\cite{Mo, Sa}),  originally developed to determine a canonical embedding of  the group of diffeomorphisms of a supermanifold into a (infinite-di\-men\-sional) supergroup.
 \end{itemize}
\begin{definition}
\label{f-supermanifolds} Let $M = (M_o, \cA_M)$ be a 
(real) supermanifold of dimension $(n|m)$ and   $\LL = \L^* W$  the exterior algebra of a fixed  vector space $W = \bR^N$.  The corresponding   {\it  $\LL$-supermanifold\/}  
is the pair  $M^{\LL} =(M_o, \cA_M^{\LL})$,  formed by the body $M_o$ of $M$ and  the sheaf 
 $$\pi^\LL: \cA_M^{\LL} = \LL \otimes_\bR \cA_M \longrightarrow M_o\ ,$$
generated by tensor products  $\h \otimes \gf$, with $\h \in \LL$ and $\gf \in \locsect{\cA_M}$. Sections of $\cA_M^{\LL}$  are called {\it $\LL$-superfunctions}
(\footnote{
A similar definition can be given for complex $\LL$-supermanifolds, where  $M = (M_o, \cA_M)$ is assumed to be complex  and  $\bK = \bR$  is replaced by   $\bK = \bC$ at all places.}).  
\end{definition}
\par
The sheaf $\cA_M^{\LL}$  is tightly related  with the sheaf of superfunctions of a particular Cartesian product of  supermanifolds (see \S \ref{cartsup}). In fact,   $\LL = \L^* W$ is isomorphic to the structure sheaf  of $\bR^{0|N} = (\{0\}, \cA_{\bR^{0|N}} = \L^* \bR^N)$  and
  $\pi^{\LL}:  \cA_M^{\LL} \longrightarrow M_o$  is naturally identifiable  with the sheaf of superfunctions $\pi: \cA_{\bR^{0|N} \times M} \longrightarrow \{0\} \times M_o \simeq M_o$ of the  supermanifold $\bR^{0|N} \times M$. 
So, we may say  that    {\it any $\LL$-supermanifold $M^{\LL} =(M_o, \cA_M^{\LL})$ is naturally identifiable with a Cartesian product of supermanifolds of the form $M^{\LL} = \bR^{0|N} \times M$}. With the help of  such identification, we may  introduce  the following fundamental objects. \par
 \medskip  
 Let  $\cU = (\cU_{o}, \cA_M{}_{\dl \cU_{o}})$ be a decomposable neighborhood and  
 $\xi = (x^i): \cU_o \longrightarrow \cU_o' \subset \bR^n$    coordinates on $\cU_o$, associated with a  system   of supercoordinates
$$\wh \xi:   \sheaf{(\L \bR^m{}^* \times \cU_o' )} \longrightarrow \cA_M{}_{\dl \cU_{o}}\ ,$$
  shortly denoted  by  $ \wh \xi = (x^i, \vartheta^\a)$. By means of the unique $\LL$-linear extension
  $$\wh \xi^\LL: \sheaf{(\L \bR^m{}^* \times \cU'_o)^{\LL}}\longrightarrow \cA^{\LL}_M{}_{\dl \cU_{o}}
$$
of $ \wh \xi = (x^i, \vartheta^\a)$,  the $\LL$-superfunctions are identified with  sums  of the form
$$ \label{f-superf}\gf = \sum_{ \begin{smallmatrix}
\a_j = 0,1 \end{smallmatrix}} \!\!\!\! \gf_{\a_1 \dots \a_m} (x^1, \dots , x^n) (\vartheta^1)^{\a_1} \wedge\dots \wedge  (\vartheta^m)^{\a_m}  \ ,$$
where $\gf_{\a_1 \dots \a_m} (x^1, \dots , x^n)$ are $\LL$-valued functions,  called  {\it components of $\gf$ in   coordinates $(x^i, \vartheta^\a)$}. 
By definitions,   $\gf$  is even  if and only  if the components $\gf_{\a_1 \dots \a_m}$, whose indices are such that  $ \sum_{i = 1}^m \a_i $ is even, are even-valued, while the other components are  odd-valued. A reversed  rule characterizes the components of odd superfunctions. \par
\medskip
A {\it morphism\/}  between $\LL$-supermanifolds $M^{\LL} \simeq \bR^{0|N} \times M$, $M'{}^{\LL} \simeq \bR^{0|N} \times M'$  is any morphism of supermanifolds  $(f, \wh f)$, in which the sheaf morphism  $\wh f:  \cA^{\LL}_{M'} \longrightarrow  \cA^{\LL}_M$ 
is $\LL$-linear, i.e. 
$\wh f(\h \otimes \gf) = \h \otimes \wh f(\gf)$ for any $\h\in\LL$.
Due to this, given supercoordinates  $(x^i, \vartheta^\a)$, $(y^j, \psi^\b)$ on $M$ and $M'$, respectively, the morphism   $(f, \wh f)$  is  completely   determined by the expressions of the $\LL$-superfunctions  
$y^j(x,  \vartheta) = \wh f(y^j) $ and $ \psi^\b(x, \vartheta) = \wh f(\psi^\b)$.
\par
\medskip
Let us  denote by   $\gJ_M^{\LL} \subset \cA_M^{\LL}$  the  $\cA_M^{\LL}$-invariant subsheaf, generated by $1 \otimes_\bR \gJ_M \subset \cA^{\LL}_M$ (see \S  \ref{firstdef} for definition of $\gJ_M$).  It can be  checked that    $\cA_M^{\LL}/\gJ^{\LL}_M$ is naturally identifiable with the sheaf  of $\LL$-valued functions $\gF^{\LL}_{M_o}$.  The  projection 
$$\epsilon: \cA^{\LL}_M  \longrightarrow  \cA^{\LL}_M/\gJ^{\LL}_M\simeq\gF^{\LL}_{M_o}$$
is  called     {\it evaluation map of  $M^{\LL}$}. For any  $\gf \in \locsect{\cA^{\LL}_M}$,  
we  use the symbols  `` $\gf|_{M_o}$'' 
to denote $\e(\gf)$ and for any $x\in M_o$ we use the symbol  `` $\gf|_{x}$''  to denote the  evaluation  $ \left.\e(\gf)\right|_x$. 
We call  {\it natural embedding of $M_o^\LL$ into $M^{\LL}$}  the morphism 
$$\imath_{M_o} = (Id_{M_o}, (\cdot)|_{M_o}): (M_o, \gF^{\LL}_{M_o}) \longrightarrow M^{\LL} = (M_o, \cA^{\LL}_M)$$
while, for any $x \in M_o$,  we call {\it natural embedding of $x$ into $M^{\LL}$}  the morphism  
 $$\imath_{x} = (Id_{x}, (\cdot)|_{x}): (\{x\}, \LL) \longrightarrow M^{\LL} = (M_o, \cA^{\LL}_M)\ .$$
  In the following,  we will use the expression   {\it supervector field of  $M^{\LL}$} to indicate   supervector fields  of   $M^{\LL} \simeq \bR^{0|N}\times M$, which  act trivially on any set of  odd supercoordinates $(\h^1, \dots,  \h^N)$ for  $\bR^{0|N}$. In a system of supercoordinates $(\h^\b, x^i, \vartheta^\a)$ of $\bR^{0|N} \times M$,  the supervector fields of $M^{\LL}$ are  derivations of the form 
 $$X = X^j \frac{\partial}{\partial x^j} + X^\a \frac{\partial}{\partial \vartheta^\a}\ ,\qquad\text{with}\ \  X^j\ ,\  X^\a \in \sect{\cA^{\LL}_{M \dl \cU_o}}\ ,$$
 i.e. with vanishing  components along the   $\frac{\partial}{\partial \h^\b}$'s. 
\par
\medskip
The sheaf over $M_o$ of  supervector fields of $M^{\LL}$ will be denoted by $\cT M^{\LL}$.  Be aware that, by definitions,  {\it $\cT M^{\LL}$ is a proper subsheaf of the tangent sheaf $\cT (M^{\LL}) = \cT ( \bR^{0|N} \times M)$} and that, as $\LL$-module, 
$$\cT M^{\LL} \simeq \cA^{\LL}_{M} \otimes_{\cA_M} \cT M\simeq\LL\otimes_{\bR}\cT M .$$ 
\par\noindent
For any $x \in M_o$ and $X\in \locsect{\cT M^{\LL}}$, we call {\it evaluation of $X$ at $x$} the map
$$
X|_{x}:\cA_M^{\LL}\longrightarrow\LL\ ,\qquad X|_{x}(\gf)\=(X(\gf))|_{x}\ .
$$
We call  {\it tangent space of $M^{\LL}$ at $x$}  the space $T_x M ^{\LL}$ generated by the evaluations at $x$  of supervector fields of $M^{\LL}$. One can check that  $T_x M ^{\LL}  = (T_{x} M^\LL)_0 + (T_{x} M^\LL)_1$,  with 
$$(T_x M ^{\LL})_\a=\left\{v:\cA_M^{\LL}\longrightarrow\LL\ , \ \ \  v(\gf\cdot\gg)=v(\gf)\cdot \gg|_{x}+(-1)^{\a |\gf|} \gf|_{x}\cdot v(\gg)\ ,\right.\phantom{aaa}
$$
$$
\left.   \phantom{aaaaaaaaaaaaaaaaa \cA_M^{\LL}} v(\gf) \in \LL_{[|\gf| + \a]_{\!\!\!\!\!\!\mod \!2}}\ \  \ \text{and}\ \  v(\eta)=0\,\text{for any}\, \eta\in\LL\right\}
$$
and that 
$T M^{\LL}|_{M_o}Ê\= \bigcup_{x \in M_o}ÊT_x M^{\LL}$ is naturally identifiable with the lambdification of $TM|_{M_o}$, i.e.  $T M^{\LL} |_{M_o}Ê\simeq \left(TM|_{M_o}\right)^\LL$. \par
\medskip
Let $\cU  = (\cU_{o}, \cA_M{}_{\dl \cU_{o}})$ be a decomposable neighborhood with  supercoordinates $\wh \xi = (x^i, \vartheta^\a)$. The sections in $\sect{\sheaf{(T M^{\LL}|_{M_o})}_{\dl \cU_o}}$   are of the form 
$$X = X^j  \left.\frac{\partial}{\partial x^j}\right|_{M_o}  +  X^\a  \left.\frac{\partial}{\partial \vartheta^\a}\right|_{M_o}\ ,\qquad X^j, X^\a \in \sect{\gF^{\LL}_{M_o}}\ ,$$
where for any $\gf\in\cA_M^{\LL}$
$$
\left.\frac{\partial}{\partial x^i}\right|_{M_o}\cdot \gf\=\left.\left(\frac{\partial}{\partial x^i}\cdot\gf\right)\right|_{M_o}\in\gF_{M_o}^{\LL}\quad ,\quad \left.\frac{\partial}{\partial \vartheta^\a}\right|_{M_o}\cdot \gf\= \left.\left(\frac{\partial}{\partial \vartheta^\a}\cdot \gf\right)\right|_{M_o}\in\gF_{M_o}^{\LL}\ .
$$
By definitions, $X$ is even (resp. odd) if and only if all $X^j$'s  are  even-valued (resp. odd-valued) and all $X^\a$'s are odd-valued (resp. even-valued). 
\par
\medskip
\subsection{Tensor fields and linear frames on $\cyrbf{L}$-supermanifolds} \hfill\par
\label{tensorfields}
For a $\LL$-supermanifold $M^{\LL} = (M_o, \cA_M^{\LL})$, we call  {\it cotangent sheaf of $M^{\LL}$}  the sheaf over $M_o$ given by 
$$ \cT^* M ^{\LL} = \Hom_{\cA^{\LL}_M}(\cT M^{\LL}, \cA^{\LL}_M)\ .$$
Local sections  of $\cT^* M^{\LL}$ are called {\it  1-forms}. A 1-form $\o$ is called  {\it  homogeneous of parity $|\o|\in\bZ_2$} if 
$$\o(\cT M^{\LL}_i)\subseteq \cA^{\LL}_M{}_{[i+|\o|]_{\!\!\!\!\!\!\mod \!2}}\qquad \text{and}\qquad \o(\gf X) = (-1)^{|\o| |\gf|} \gf \o(X)$$
for any homogeneous $\gf \in \locsect{\cA^{\LL}_M}$,  $X \in \locsect{\cT M^{\LL}}$. As for  usual supermanifolds, we call   {\it full tensor sheaf of $M^{\LL}$}  the sheaf $  \otimes_{\cA^{\LL}_M}\!\!\!\! < \cT M^{\LL}, \cT^* M^{\LL}> $, 
  generated by  tensor products ($\bZ_2$-graded over $\cA^{\LL}_M$) of $\cT M^{\LL}$ and $\cT^* M^{\LL}$ (see \S \ref{supervect}).  A local section of $\otimes_{\cA_M^{\LL}}\!\!\!\! < \cT M^{\LL}, \cT^* M^{\LL}> $ is  called {\it tensor field of type $(p,q)$} if  it   is sum of tensor products of $p$ vector fields and $q$ 1-forms. 
  The notions of ``homogeneity'' and ``parity'' of tensor fields on $\LL$-supermanifolds are defined in full analogy with  the corresponding notions   on supermanifolds.  Also the  definitions of skew-symmetric tensors of type $(p,0)$,  $q$-forms, exterior differential and interior multiplication with vector fields are defined exactly as for usual supermanifolds.  See  \S  \ref{p-forms} for a  brief review of all such notions and    \cite{Ko} for detailed definitions. \par
\medskip
Given a  1-form $\o\in\locsect{\cT^* M^{\LL}}$, we call {\it evaluation of $\o$ at $M_o$} the $\LL$-linear bundle morphism
$$\o|_{M_o}: TM^{\LL}|_{M_o}Ê\longrightarrow M^{\LL}_o\ ,\qquad \o|_{M_o}(X) = \left.\o(\wh X)\right|_{M_o}$$
for any $X \in \locsect{TM^{\LL}|_{M_o}}$ of the form  $X = \wh X|_{M_o}$ for some $\wh X \in \sect{\cT M^{\LL}}$. 
This definition naturally extends  to  all other   tensor fields.\par
For  $x \in M_o$ and $\o\in \locsect{\cT^* M^{\LL}}$,  the {\it evaluation of $\o$ at $x$} is the $\LL$-linear map
$$
\o|_{x}:T_x M ^{\LL}\longrightarrow\LL,\qquad \o|_{x}(X)\=(\o(\wh X))|_{x}\ ,\quad \text{with}\ X = \wh X|_{x}\ .
$$
We call  {\it cotangent space of $M^{\LL}$ at $x$}  the space $T^*_x M ^{\LL}$ generated by the evaluations at $x$  of 1-forms of $M^{\LL}$. One can check that  $T^*_{x} M^\LL= (T^*_{x} M^\LL)_0 + (T^*_{x} M^\LL)_1$ with
$$
\left(T_x^* M ^{\LL}\right)_\a\!\!\!\! =\!\left\{\o:T_x M ^{\LL}\rightarrow\LL,\o(\eta v)=(-1)^{\a |\eta|}\eta\o(v),\  \o(v) \in \LL_{[|v| + \a]_{\!\!\!\!\!\!\mod \!2}},\right.$$
$$\left.   \phantom{aaaaaaaaaaaaaaaaaaaaaaaaaaaaaaaaaaaaaaaaaa \cA_M^{\LL}}  \eta\in\LL,\, v\in T_x M ^{\LL} \right\}\!\!.
$$

As for $TM^{\LL}|_{M_o}$, one can check that   $T^* M^{\LL}|_{M_o} = \bigcup_{x \in M_o}ÊT^*_x M^\LL$ is identifiable with $\left(T^* M|_{M_o}\right)^\LL$.
It follows that 
$$\textstyle\left(\bigotimes^p T M^{\LL}|_{M_o}\right)\bigotimes \left(\bigotimes^qT^* M^{\LL}|_{M_o}\right) \simeq
\left(\left(\bigotimes^p T M|_{M_o}\right)\bigotimes \left(\bigotimes^qT^* M|_{M_o}\right)\right)^\LL\!\!.
$$
\par
\bigskip
In order to get short statements, close  to familiar  sentences on  smooth manifolds,   we  consider the following definitions.  Given an open subset  $\cU_o \subset M_o$, we  call  {\it open subset of $M^\LL$} the  $\LL$-supermanifold  $\cU^\LL = (\cU_{o},  \cA^\LL_M{}_{\dl \cU_{o}})$ and,  if  $x_o \in \cU_o$, we say that {\it $\cU^\LL$ is a neighborhood of $x_o$ in $M^\LL$}. Moreover: 
\begin{definition} 
Let  $\cU^\LL$ be an open subset of $M^\LL$.
An ordered set  $(e_1, \dots, e_{n + m})$ of supervector fields of $\cU^\LL$  is called {\it (local) frame field}  if: 
\begin{itemize}
\item[i)] it is a collection of $ \cA^\LL_M{}_{\dl \cU_{o}}$-linearly independent generators for the $ \cA^\LL_M{}_{\dl \cU_{o}}$-module $\sect{\cT \cU^\LL}$; 
\item[ii)] all  $e_i$'s, with $1 \leq i \leq n$, are even, while the $e_j$'s, with $n+1 \leq j \leq m + n$, are odd. 
\end{itemize}
\end{definition}
For any open $\cU_o \subset M_o$,  we denote by  $\cF r^\LL(\cU)$  the collection of linear frames on the corresponding open subset $\cU^\LL \subset M^\LL$. The sheaf $\pi: \cF r^\LL(M) \longrightarrow M_o$,  determined by  the pre-sheaf $\left\{\cU_o \longrightarrow \cF r^\LL(\cU)\right\}$, is called   {\it sheaf of frame fields of $M^\LL$} (see also \cite{ACDS}). 
Notice that,  for any local frame field $(e_1, \dots, e_{n+m})$, the  evaluations $(e_1|_{M_o}, \dots, e_{n+m}|_{M_o})$ constitute a local frame field for the graded vector bundle $TM^\LL|_{M_o} \simeq (TM|_{M_o})^\LL$. 
\medskip
\subsection{Flows,  Lie derivatives and linear connections}\hfill\par
Given a    $\LL$-supermanifold $M^\LL = (M_o, \cA^\LL_M)$, we call {\it (smooth) 1-parameter group of automorphisms}  any  morphism of supermanifolds (here, we think of   $M^\LL $ as a  Cartesian product of supermanifolds; see \S \ref{section2.2})
$$(\Phi, \wh \Phi): ]-a,a[ \times M^\LL  \longrightarrow M^\LL\ ,\qquad a \in \bR\cup \{\infty\}$$
such that: 
\begin{itemize}
\item[i)] for any $t \in ]-a,a[$, the morphism $(\Phi_t, \wh \Phi_t): M^\LL \longrightarrow M^\LL$ defined by 
$$\Phi_t \= \Phi(t, \cdot)\ \ \text{and}\ \ \wh \Phi_t: \cA_M^{\LL} \longrightarrow\Phi_t{}_*(\cA_M^{\LL})\ ,\ \  \wh \Phi_t( \gf) \= \e_t(\wh\Phi(\gf))\ ,$$ 
is an isomorphism of $\LL$-supermanifolds; 
\item[ii)]Ê$(\Phi_0, \wh \Phi_0) = \Id_M$ and   $(\Phi_{t}, \wh \Phi_{t}) \circ (\Phi_{s}, \wh \Phi_{s}) = (\Phi_{t+s}, \wh \Phi_{t+s}) $ for any $t, s$ such that $t + s \in ]-a, a[$.
\end{itemize}
The {\it supervector field of 
$(\Phi, \wh \Phi)$} is the derivation $V\in\sect{\cT M^\LL}$  defined by 
$$ \label{3.1}ÊV \cdot \gf = \lim_{h \to 0} \frac{1}{h} \left(\wh \Phi_h(\gf) - \gf\right) = \left.\frac{d \wh \Phi_t(\gf)}{dt}\right|_{t= 0}$$
and we say that {\it $(\Phi, \wh \Phi)$ is a  flow of $V$}. 
In  analogy with  smooth manifolds, it is possible to define {\it local} 1-parameter groups of automorphisms and corresponding supervector fields. We leave to the reader the task of guessing  the appropriate definitions.
\par
\medskip
One can directly check that the supervector field of 
a local 1-parameter group of automorphisms is always  {\it even}. The converse is also true. In fact:
\begin{lem}[\cite{MM}, Thm. 4] \label{lemma3.4} For any  even $V \in \locsect{\cT M^\LL}$, there exists  a local 1-parameter group of automorphisms $(\Phi, \wh \Phi)$, which is a flow of $V$. Given two flows of $V$,  defined on open subsets $I\times\cU^\LL $, $I'\times\cU^{\LL}$, the isomorphisms $\wh \Phi_t$, $\wh \Phi'_t$ coincide for  any $t  \in I \cap I'$. 
\end{lem}
For an even supervector field $V$, we   denote by $\Phi^V_t$ the corresponding flow.
One can check that all    properties of  local flows on smooth manifolds have   analogues  for   local flows on $\LL$-supermanifolds. 
In particular, for any even supervector field $V$, the map $\a \longmapsto \Phi^{V}_{-t}{}_*(\a)$  on tensor fields $\a$  of  $M^\LL$ is such that
$$ \left. \frac{d \Phi^{V}_{-t}{}_*(\a)}{dt}\right|_{t = 0} = \cL_V \a\ ,$$
where  ``$\cL_V(\cdot)$''  denotes  the  unique derivation of tensors,  which is compatible with contractions and  such that $\cL_V (\gf) = V \cdot \gf$,  $\cL_V X = [V,X]$  for any  superfunction $\gf$ and supervector field $X$. 
We call  $\cL_V\a $ the  {\it Lie derivative of $\a$ along $V$}. One can  check that, writing tensor fields w.r.t.
frames $\left(\frac{\partial}{\partial x^i}, \frac{\partial}{\partial \vartheta^\a}\right)$ and coframes $\left(d x^i, d \vartheta^\a\right)$,  the expression for    $\cL_V \a$ is identical to  the  usual formula for Lie derivatives on smooth manifolds.
\par
\begin{definition}
A {\it linear connection}  on $M^{\LL}$ is a linear, even  map of sheaves of $\LL$-moduli
$$\nabla: \cT M^{\LL}\otimes_{\bR}\cT M^{\LL}\longrightarrow \cT M^{\LL}$$
such that
\begin{itemize}
\item[1)]
$\nabla_{\gf X}Y=\gf \nabla_{X}Y$,
\item[2)]$\nabla_{X}\gf Y=(X\gf)Y+(-1)^{|X||\gf|}\gf\nabla_{X}Y$
\end{itemize}
for any homogeneous $\gf\in \mathcal{A}_{M}^{\LL}$, $X, Y\in\cT M^{\LL}$.
The {\it torsion} $T$ and the {\it curvature}Ê $R$ of a linear connection $\n$ are the  tensor fields of type  $(1,2)$ and $(1, 3)$, respectively,   such that,  for any homogeneous supervector fields $X$, $Y$, $Z \in \sect{\cT M^\LL}$, 
$$ T_{X Y} \= \n_X Y - (-1)^{|X||Y|}\n_Y X - [ X, Y]\,\, ,$$
$$ R_{X Y} Z \= \n_X \n_Y  Z - (-1)^{|X||Y|}\n_Y \n_X Z - \n_{[ X, Y]} Z\  .$$
\end{definition}
Given a linear connection $\nabla$,  the {\it  induced connection on the vector bundle $\pi:T M^{\LL}|_{M_o}\longrightarrow M_o$} is the map
$$
\n:\sect{(TM_o)^{\LL}}\times\sect{T M^{\LL}|_{M_o}}\longrightarrow\sect{T M^{\LL}|_{M_o}}\ ,\ \n_X Y \= (\n_{\wh X} \wh Y)|_{M_o}\ , $$
where
$\wh X$, $\wh Y \in \locsect{\cT M^{\LL}}$ are such that
$X= \wh X|_{M_o}$, $Y = \wh Y|_{M_o}$.\par
\bigskip
 \section{Super-spacetimes and supergravities  on $\cyrbf{L}$-supermanifolds}
\setcounter{equation}{0}
\subsection{Admissible super Poincar\`e algebras}\hfill\par
Given a flat pseudo-Riemannian space  $V = \bR^{p,q}$, with the scalar product $< \cdot, \cdot>$ and Poincar\`e algebra    $\gp(V) = Lie(Iso(\bR^{p,q})) = \so(V) + V$,  the   {\it super-extensions of $\gp(V)$} are the Lie superalgebras   $\gg = \gg_0 + \gg_1$  with the following properties:
\begin{itemize}
\item[a)] $\gg_0 = \gp(V) = \so(V) + V$;
\item[b)] $\gg_1=S$ is   an irreducible   spinor module (i.e. an irreducible real representation of the Clifford algebra $\cC \ell(V)$ of $V$) and the adjoint action $\ad_{\so(V)}|_S:  S \longrightarrow S$ coincides with the standard action of $\so(V)$ on $S$ (i.e. $[A,s]  = A \cdot s$ for any $A \in \so(V)$, $s \in S$);
\item[c)]  $[V, S] = 0$ and  $[S, S] \subseteq V$.
\end{itemize}
Any super-extension of  $\gp(V)$ is  called    {\it super Poincar\`e algebra}. \par
\smallskip
We recall that, given an irreducible spinor  module $S$, any  super-extension  $\gg = (\so(V) + V) + S$  is uniquely determined by the $\so(V)$-invariant tensor
$L \in  \vee^2 S^* \otimes V$  defining  the Lie bracket $[\cdot ,\cdot ]|_{S \times S}: S \times S\longrightarrow V$ and any  $\so(V)$-invariant tensor  of this kind  gives a super-extension of $\gp(V)$. \par
\medskip
A tensor   $L \in \vee^2 S^* \otimes V$  is called {\it admissible} if the  associated tensor
$L^*  \in \vee^2 S^* \otimes V^*$, defined by $L^*(s, s', v) \=  <L(s,s'), v>$, is of the form
$$\label{admissible}L^*(s,s', v) = \beta(v\cdot s, s')  $$
 (here ``$\cdot$''   stands for  Clifford product) for some  non-degenerate $\so(V)$-invariant bilinear form  $\b$  on $S$  such that:
\begin{itemize}
\item[1)] it is  either symmetric or skew-symmetric;
\item[2)] the Clifford multiplications $ v \cdot ( \cdot): S \longrightarrow S$, $v \in V$,  are either  all $\b$-symmetric or all $\b$-skew-symmetric;
\item[3)] if $S$ decomposes into irreducible $\so(V)$-moduli $S = S^+ + S^-$, then $S^\pm$ are either mutually $\b$-orthogonal or both $\b$-isotropic.
\end{itemize}
It is known that  {\it any admissible tensor  is $\so(V)$-invariant, it corresponds to a  super Poincar\`e  algebra 
and  there is a basis for   $(\vee^2 S^* \otimes V)^{\so(V)}$
made of   admissible elements} (\cite{AC}). \par
\medskip
  A   super Poincar\`e algebra,  determined   by an admissible tensor,  is called  {\it admissible}. In this case,  $V +S$ is naturally
endowed with the non-degenerate  bilinear form $(\cdot, \cdot)$, called {\it extended inner product}, defined by
$$ \label{innerproduct} (\cdot, \cdot)|_{V \times S} = 0\ ,\quad (\cdot, \cdot)|_{V \times V} =  <\cdot ,\cdot >\ ,\quad (\cdot, \cdot)|_{S \times S} =\b\ .$$
From now on,   {\it any super Poincar\`e algebra  is assumed to be admissible and $(\cdot, \cdot)$  always indicates  the  extended inner product on $V + S$}.
\par
\medskip
\subsection{Distributions, Levi forms   and  super-spacetimes of type $\gg$}\hfill\par
\begin{definition} A {\it distribution} $\cD$ on  a $\LL$-supermanifold  $M^{\LL}$ is a $\bZ_2$-graded subsheaf of $\cA_{M}^\LL$-modules of $\cT M^{\LL}$,  which is locally a direct factor in  $\cT M^{\LL}$.  It is  called {\it odd} (resp. {\it even}) {\it of rank q}  if for any $x_o \in M_o$, there  exists neighborhood $\cU^\LL$ of $x_o$ in $M^\LL$ and a  local frame field $(e_i)$ on $\cU^\LL$,  such that $q$ of its  odd (resp. even) elements are generators for   $\cD_{\dl \cU_o}$. \par
  The {\it Levi form} of a distribution $\cD$ is the sheaf morphism
\beq \label{leviform} {\cL}: \cD \times \cD \longrightarrow \cT M^\LL /\cD\ ,\eeq
determined by    the  map between local vector fields
$$\cL: \locsect {\cD} \times \locsect {\cD} \longrightarrow \locsect{\cT M^\LL/\cD}\ ,\ \   \cL(X, Y) \= [X,Y]/\locsect{{\cD}}\ .$$
\end{definition}
\medskip
Notice that, if  a complementary distribution $\cD^\perp$ of $\cD$ is fixed  (possibly only locally defined),   the  quotient map  $p: \cT M^\LL \longrightarrow \cT M^\LL/\cD$ gives a sheaf  isomorphism $p|_{\cD^\perp}: \cD^\perp \overset{\simeq} \longrightarrow \cT M^\LL/\cD$ and  the Levi form \eqref{leviform} is completely  determined by  the associated  tensor field $\cL^{(\cD^\perp)}$   of type $(1,2)$ defined by 
$$\label{levitensor}  \cL^{(\cD^\perp)}(X, Y) \= (p|_{\cD^\perp})^{-1}\circ\left(\cL(\pi^\cD(X), \pi^\cD(Y))\right)\qquad X, Y \in \locsect{\cT M^\LL}\ ,$$
where $\pi^\cD: \cT M^\LL \longrightarrow \cD$, $\pi^{\cD^\perp}: \cT M^\LL \longrightarrow \cD^\perp$ are  the standard projections,  determined by the   decomposition $\cT M^\LL = \cD \oplus \cD^\perp$. We call $\cL^{(\cD^\perp)}$ the {\it  Levi tensor  of   $\cD$ determined by  $\cD^\perp$}. 
Whenever the context makes  clear  which complementary distribution is considered,  the  symbol $\cL$ will be used to indifferently  denote the Levi form and the Levi tensor. \par
\bigskip
Consider now a super Poincar\`e algebra $\gg = (\so(V) + V) + S$ and a corresponding connected  homogeneous
superspace $G/H = (G_o/H, \cA_{G/H})$,  with   $Lie(G_o) =   \so(V) + V$ and $H \subset G_o$  connected subgroup with $ \gh = Lie(H) = \so(V)$. 
 We call {\it flat super-spacetime of type $\gg$\/}
 any $\LL$-supermanifold $M^\LL$ associated with $M = G/H$. \par
 \medskip
Since the subspaces $V$, $S \subset \gg$ are $\Ad_H$-invariant, the homogeneous superspace $M = G/H$ and the $\LL$-supermanifold  $M^\LL$ admit  the complementary $G$-invariant distributions $\cD^{\gg}$, $\cD^{\gg}{}^\perp$ defined as follows.\par
Let $\cS$, $\cV \subset \cT G$ be the distributions of $G$, generated by the left-invariant vector fields  in $S$ and $V$,  respectively, and denote by 
$$(\pi_o, \wh \pi): G = (G_o, \cA_G) \longrightarrow G/H = (G_o/H, \cA_{G/H})$$
the natural superspace morphism from $G$ onto $G/H$. Since $S$, $V\subset \gg$ are $\Ad_H$-invariant, the distributions $\cS$,
$\cV$ are invariant under the right action of $H$ and hence locally generated by  vector fields  $s_\a \in \cS$,  $v_i \in \cV$  {\it invariant under the right-action of $H$}. For any such vector field,  there exists a  unique  vector field $\wt s_\a$  or $\wt v_i$ in  $\cT G/H$ such that
$$s_\a \cdot \wh \pi(\gf)  = \wh \pi (\wt s_\a \cdot \gf)\quad\text{or}\quad v_i \cdot \wh \pi(\gf) = \wh \pi (\wt v_i \cdot \gf)\qquad \text{for any} \ \gf \in \locsect{\cA_{G/H}}\ .$$
The vector fields $\wt s_\a$ and $\wt v_i$ generate two complementary $G$-invariant distributions in $\cT G/H$ (and in $\cT G/H^\LL$), which we call 
$\cD^\gg$ and  $\cD^\gg{}^\perp$, respectively. \par
\smallskip
Let $W = (V , \cA_W)$ denote   the connected super-subgroup of $G$, with  associated sHC-pair $(V , V+S)$ (see \S \ref{appendixA2}). Given two     bases $(e_i)_{i = 1, \dots n}$, $(e_\a)_{\a = 1, \dots, m}$, for $V$ and $S$  respectively, denote by $(E_i, E_\a)$ the local frame field on $M^\LL =   (G_o/H, \cA_{G/H}^\LL)$, formed by the  $W$-invariant  even and odd vector fields with 
$$E_i|_o = e_i \ ,\qquad E_\a|_o = e_\a\ ,\qquad \text{where}\ \ o = e H \in G_o/H\ .$$
One can directly check that: 
\begin{itemize}
\item[a)]  the  $E_i$'s are even generators for $\cD^\gg{}^\perp$; 
\item[b)]  the  $E_\a$'s are odd generators for $\cD^\gg$; 
\item[c)]   the Levi tensor $\cL^\gg$ of $\cD^\gg$ (determined by $\cD^\gg{}^\perp$) is such that
\beq\label{levigform}Ê \cL(E_i, E_j) = 0\ ,\qquad \cL(E_i, E_\a) = 0 \ ,\qquad \cL(E_\a, E_b) = \cL^i_{\a \b}E_i\ ,\eeq
where  $\cL^i_{\a \b}$ are the  structure constants defined by  $[e_\a, e_\b] = \cL^i_{\a \b} e_i$.
\end{itemize}
\par
Now, consider a  super Poincar\`e algebra $\gg = (\so(V) + V) + S$ and  let   $n = \dim V$,  $m = \dim S$. \par
\begin{definition}
\label{superspacetime1}
A  {\it super-spacetime of type $\gg$\/} is a triple $(M^{\LL}, M_o^{\LL},  \cD)$, where: 
\begin{itemize}
\item[--] $M^{\LL} = (M_o,\cA_M^{\LL})$ is  $\LL$-supermanifold   of dimension $(n|m)$;
\item[--] $M_o^{\LL}=\LL\times M_o $ where $M_o$ is the body of $M^{\LL}$; 
\item[--] $\cD \subset \cT M^{\LL}$ is an odd distribution of rank $m$ satisfying  the following  ``uniformity assumption'':
  for any $x_o\in M_o$,   there is a neighborhood $\cU^\LL$ of $x_o$  in $M^\LL$,  a local frame field $(E_i, E_\a)$ on $\cU^\LL$ and an associated  basis $(e_i, e_\a)$ for $V + S$ such that  
 \begin{itemize}
 \item[i)] the odd fields $E_\a$ generate $\cD$ and the even fields $E_i$ generate a complementary distribution $\cD^\perp$; 
 \item[ii)] the Levi tensor $\cL$ of $\cD$, determined by $\cD^\perp$,  is  
such that $\cL(E_\a, E_\b) = \cL^i_{\a \b}E_i$, where $\cL^i_{\a \b}$ are the constants   in \eqref{levigform}; 
\item[iii)]  if   $S=S^+ + S^-$  is sum of irreducible $\so(V)$-moduli, the basis $(e_\a)$ is formed by two  bases $(e_\b)$, $(e_{\dot \b})$ for $S^+$,  $S^-$,   respectively,  and the corresponding fields $E_\b$, $E_{\dot \b}$ generate   complementary subdistributions $\cD^{\pm} \subset \cD$.
\end{itemize}
\end{itemize}
\end{definition}
\medskip
\subsection{Supergravities of type $\gg$}\hfill\par
\begin{definition}\label{supergravityfields2}
A {\it gravity field\/}  on a super-spacetime $(M^{\LL}, M_o^\LL, \cD)$ is a pair $(g, \n)$, where $g$ is an even tensor field of type $(0,2)$ and $\n$ is a linear connection on $M^{\LL}$   such that 
\begin{itemize}
\item[i)] the tensor $g$ is such that,  for any $x_o\in M_o$,   there is neighborhood  $\cU^\LL$ of $x_o$ in  $M^\LL$,  a local frame field $(E_A) = (E_i, E_\a)$ on $\cU$ and an associated  basis $(e_A) = (e_i, e_\a)$ for $V + S$ such that: 
 \begin{itemize}
  \item[a)] the odd fields $E_\a$ generate $\cD$, while  the even fields $E_i$ generate  a complementary  distribution $\cD^\perp$; if $S=S^+ + S^-$, the   $e_\a$'s  are given by   bases $(e_\b)$, $(e_{\dot \b})$ for $S^+$, $S^-$  respectively, and the  corresponding fields $E_\b$, $E_{\dot \b}$ generate   complementary subdistributions $\cD^{\pm} \subset \cD$; 
 \item[b)]   the fields $E_A$ are such that  $g(E_A, E_B) \equiv  (e_A, e_B)$, where $(\cdot, \cdot)$ denotes the extended inner product of $V + S$; 
 \item[c)]Êthe components in the frame field $(E_A)$ of the Levi tensor $\cL$ of $\cD$,  determined by $\cD^\perp$,  are constant and equal to those in \eqref{levigform}. 
 \end{itemize}
 \item[ii)]  the distribution $\cD$ is $\n$-stable and, if $S = S^+ + S^-$, both distributions $\cD^\pm$ are $\n$-stable; 
\item[iii)]   $\n g = 0$ and  $\n \cL = 0$. 
\end{itemize}
A {\it supergravity of type $\gg$\/} is a  pair  $\cG= ((M^\LL, M_o^\LL, \cD),(g,\n))$ formed by  a  super-spacetime $(M^\LL, M_o^\LL, \cD)$ of type $\gg$  and a gravity field $(g,\n)$  on it.
\end{definition}
For a given  supergravity $\cG= ((M^\LL, M_o^\LL, \cD),(g,\n))$, we call
{\it spinor bundle}  the  fermionic subbundle of $T M^\LL|_{M_o}$,  given  by 
$\cS:=\cD|_{M_o}$,  and we call {\it physical fields} the following
objects:
 \begin{itemize}
  \item[--] the field (=even section) $\vartheta$  in the fermionic bundle $T^*M_o  \otimes_{M_o}\cS$ over $M_o$, called  {\it gravitino\/}, defined by
$$
\label{gravitino}
\vartheta(X) \= \pi^{\cD}(\wh X)|_{M_o}\qquad  \text{for any}\  X \in \gX_{\text{loc}}(M_o)\ ,$$
where $\pi^\cD: \cT M^\LL = \cD \oplus \cD^\perp \longrightarrow \cD$ denotes the natural projection onto $\cD$   and $\wh X$ any field in $\locsect{\cT M^\LL}$  with $\wh X|_{M_o}Ê= X$; 
\item[--]  the  field $\wh g$ in the bosonic bundle $(\vee^2 T^* M_o)^{\LL}$,  called  {\it graviton\/}, defined by
$$
\label{graviton}
\phantom{aaaaaa} \wh g(X, Y) = \left.g(\pi^{\cD^{\perp}}(\wh X),\pi^{\cD^{\perp}}(\wh Y) )\right|_{M_o}\ \  \text{for any}\  X, Y\in \gX_{\text{loc}}(M_o)\ ,$$
where $\pi^{\cD^\perp}: \cT M^\LL = \cD \oplus \cD^\perp \longrightarrow \cD^\perp$ is the natural projection onto $\cD^\perp$ and $\wh X, \wh Y$ are fields in  $\locsect{\cT M^\LL}$   with  $\wh X|_{M_o}Ê= X$, $\wh Y|_{M_o}Ê= Y$;
\item[--] the field $\bA$ in the bosonic bundle  $ T^*M_o \otimes_{M_o} \cS^* \otimes_{M_o}\cS$, called  {\it A-field},
defined by
$$
\phantom{aaaaaa} \bA_{X s} \=   -\left. \pi^{\cD} (T_{\wh X\wh s})\right|_{M_o}\qquad \text{for any}\ X\in \gX_{\text{loc}}(M_o)\ ,\  s \in \locsect{\cS}\ ,$$
where $T$ is the torsion of $\n$ and $\wh X, \wh s$ are fields in  $\locsect{\cT M^\LL}$  such that $\wh X|_{M_o}Ê= X$, $\wh s|_{M_o} = s$; 
 \item[--] the connection $D: \gX(M_o) \times \gX(M_o) \longrightarrow  \sect{(TM_o)^{\LL}}$, called  {\it metric connection\/}, defined by
$$ \label{metconn}  D_X Y \= (\left.\pi^{\cD^{\perp}}\right|_{\sect{(TM_o)^{\LL}}})^{-1}\left.\left(\n_{\wh X}\left( \pi^{\cD^{\perp}}(\wh Y)\right)\right)\right|_{M_o}$$
where $\wh X, \wh Y$ are fields in $\locsect{\cT M^\LL}$  such that $\wh X|_{M_o}Ê= X$, $\wh Y|_{M_o}Ê= Y$; 
\item[--]   the  connection $\bD: \gX(M_o) \times \G(\cS) \longrightarrow \G(\cS)$,  called {\it spinor connection\/},  defined by
$$
\label{spinorconn}
\bD_X s \=  \left.\n_{\wh X} {\wh s}\right|_{M_o} + \bA_X s\ ,$$
where $\wh X, \wh s$ are fields in  $\locsect{\cT M^\LL}$  such that $\wh X|_{M_o}Ê= X$, $\wh s|_{M_o} = s$; 
\end{itemize}
\par
\medskip
Notice that  the values $\wh g|_{x}$,  $x \in M_o$,  of the graviton are identifiable with $\LL$-bilinear maps  $\wh g|_x: (T_{x}M_o)^\LL\times (T_{x}M_o)^\LL \longrightarrow \LL$ and not with classical  scalar products of the tangent spaces $T_x M_o$, as in \cite{SaS}. However, $\wh g$ is a bosonic field  and its components $\wh g_{ij} = \wh g\left(\frac{\partial}{\partial x^i},   \frac{\partial}{\partial x^j}\right)$ in any coordinate frame $\frac{\partial}{\partial x^i}$ of $M_o$,  are even-valued $\LL$-functions,  which behave as components of a pseudo-Riemannian  metric. Moreover, the tensor field $\wh g ^\bR = \wh g_{ij}^\bR d x^i \otimes dx^j$, determined  by the $\bR$-valued parts $\wh g_{ij}^\bR$ of the maps $\wh g_{ij}: \cU_o \subset M_o\longrightarrow \LL = \bR + W + \L^2 W + \dots$,  is   a    pseudo-Riemannian metric of signature $(p,q)$ in the usual sense.   Notice also that, as in \cite{SaS},  {\it  the (real part of the) metric connection $D$  is  a metric connection for $\wh g^\bR$}. 
\medskip
\subsection{The Principle of General Covariance on $\LLbis$-supermanifolds}\hfill\par
\label{generalcovariance}
From \S \ref{section2}, we know  that automorphisms, even tensor fields and Lie derivatives by  even vector fields of a  $\LL$-supermanifold  behave as diffeomorphisms,  tensor fields and Lie derivatives of a smooth manifold $M$, endowed with  a  distinguished  submanifold $M_o$ and  a sheaf of  $\LL$-valued functions. Moreover, modulo simple adjustments of signs, their expressions in supercoordinates  are identical to those  on smooth manifolds. This explains why   {\it the ``simple-minded  approach'',  which (locally) deals with  supermanifolds as  smooth spaces, with  points  labeled by bosonic and fermionic coordinates} (see e.g. \cite{CDF}, p. 338),    {\it brings in fact to correct conclusions}. \par
\smallskip
Due to this, we   may safely claim that the  results on  supergravities, stated in \cite{SaS}   for smooth manifolds,  hold for supergravities on $\LL$-supermanifolds as well.  We can also  re-formulate the Principle of General Covariance of  that paper   in terms of $\LL$-supermanifolds as follows.\par
\smallskip
First of all, notice that for a super\-gravity $\cG=$  $((M^\LL, M_o, \cD)$, $(g, \n))$    of type $\gg$,  any (local) automorphism   $ \varphi:   M^\LL \rightarrow M^\LL$
determines the new supergravity  of type $\gg$
\beq \label{actionofdiffeomorphisms} \cG' =  \varphi_*(\cG) \= ((M^\LLbis, M_o^\LL, \varphi_*(\cD)),((\varphi^{-1})^{*} g, ( \varphi^{-1})^{*}\n))\ .\eeq 
The principle   can be now stated as follows. \par
\smallskip
\moveleft 0.6cm
\vbox{\begin{itemize}
\item[]
  {\it A collection $\cE_o$ of constraints and equations  on physical fields of  supergravities of type $\gg$ satisfies the  {\rm Generalized Principle of  Infinitesimal General  Covariance} if:
\begin{itemize}
\item[i)] there is a system $\cE$ of constraints and equations on $(\cD,g,\n)$, such that any (local) solution of $\cE$ determines physical fields which solve $\cE_o$,  and every (local) solution of $\cE_o$ can be obtained in this way;
\item[ii)] the class of (local) solutions of $\cE_o$ is invariant under all actions \eqref{actionofdiffeomorphisms}, where $\cG$ is given by a solution of $\cE$ and
$\varphi = \Phi^X_t$  is a flow of an  even supervector field $X \in \locsect{\cT M^{\LL}}$.
\end{itemize}
}
\item[]
\item[]  {\it The system  $\cE_o$ is called {\rm manifestly covariant} if there exist a system $\cE$, which satisfies  (i)  and is of tensorial type.
}
\end{itemize}
}
\noindent A manifestly covariant  system $\cE_o$ {\it automatically satisfies  the Gen\-er\-al\-ized
Prin\-ci\-ple of General Covariance} (\cite{SaS}).
\par
\bigskip
\section{Supergravity in 11 dimensions}
\setcounter{equation}{0}
\subsection{Notation}
\label{notation}
\hfill\par
Let  $\gg =   \so(V) + V + S$ be a super Poincar\`e algebra with $(V, <,>) = \bR^{p,q}$, $n = p + q$. We always assume that: 
\begin{itemize}
\item[--] $(e_A) = (e_i, e_\a)$ is  a fixed basis for $V + S$, with $ (e_i)$ orthonormal basis of $(V, <,>)$, 
i.e.   $< e_i, e_j> = \e_i \d_{ij}$ with  $ \e_i=$  $ \left\{\smallmatrix \phantom{-}1 & \ \ & \text{if}\ 1& \leq i \leq p\ ,\\ - 1 & \ \ & \text{if}\ p+1 &\leq i \leq n\ ; \endsmallmatrix\right.$
\item[--] $(e^A) = (e^i,e^\a)$ is the  dual  basis  for  $V^* + S^*$ and 
$\o_o  = Êe^1 \wedge  \ldots \wedge e^n$;    we use the notation $ \e_{j_1 \ldots j_n} \= \o_o(e_{j_1}, \ldots, e_{j_n})$; 
\item[--] $(\cdot)^\sharp: \bigotimes^{r} V^*  \longrightarrow    \bigotimes^{r} V $ is the isomorphism induced by 
the duality  map $(\cdot)^\sharp: V^*\longrightarrow V$ determined  by the relation  $< \a^\sharp, v> = \a(v)$; 
\item[--]   $\ast: \Lambda^r V^* \longrightarrow \Lambda^{n-r} V^*$ is the Hodge star operator,  determined by  $\o_o $, i.e.  if   
$\displaystyle \a = \!\!\!\!\!\!\!\sum_{j_1 < \ldots < j_r} \a_{j_1 \ldots j_r}\ e^{j_1} \wedge  \ldots \wedge e^{j_{r}}$ and $\displaystyle \a^\sharp = \!\!\!\!\!\! \sum_{m_1 < \ldots < m_r} \a^{m_1 \ldots m_r}\ e_{m_1} \wedge  \ldots \wedge e_{m_r}$,  
then 
$$\ast \a =  \frac{1}{r!(n-r)!} 
\e_{m_1 \ldots m_{r} j_1 \ldots j_{n-r}}  \alpha^{m_1 \ldots m_r} \ e^{j_1} \wedge  \ldots \wedge e^{j_{n-r}}\ ;$$
\item[--] $(M^{\LL}, M_o^{\LL},  \cD)$ is a  super-spacetime   of type $\gg$; 
\item[--]  $g$ is a tensor field  of type $(0,2)$ on $M^\LL$, satisfying Definition \ref{supergravityfields2} (i) and  $\cD^\perp$ is the corresponding complementary distribution; 
\item[--] $(E_A) = (E_i, E_\a)$  is  a frame field  on an  open subset $\cU^\LL$ of $M^\LL$, 
 associated with  $(e_A)$ and satisfying  Definition \ref{supergravityfields2}  (i);  $(E^A) = (E^i, E^\a)$ is   the  dual coframe field.
\end{itemize}
For a fixed choice of  $g$,  we denote by 
$(\cdot)^\sharp:  \bigotimes^{r} \cT^* M^\LL  \longrightarrow  \bigotimes^{r}  \cT M^\LL $
 the isomorphism induced by the duality
 $$(\cdot)^\sharp: \cT^* M^\LL \longrightarrow \cT M^\LL\quad\text{such that}\quad  g(\a^\sharp, X) = \a(X)$$
 for any  $X \in \locsect{\cT M}$, $\a \in \locsect{\cT^*M}$. If   $\a = \a_A E^A$,   then  $ \a^\sharp = (\h^{AB} \a_B) E_A$ where  $[\h^{AB}] = \left[\h_{AB}\right]^{-1}$ with $\h_{AB} \= (e_A, e_B)$. \par
\medskip
\subsection{Clifford products}
\label{products}
\hfill\par
 Let  $\cB = (e_i)$ be an orthonormal basis for  $V = \bR^{p,q}$, 
and,   given a spinor representation  $\cdot : \cC\ell(V)\times S \longrightarrow S$ of the Clifford algebra $\cC \ell(V) = \cC\ell_{p,q}$ onto   $S=\bK^N$, $\bK = \bR$, $\bC$, $\bH$,  let us  denote by $\G_i \in \ggl_N(\bK)$, $1 \leq i \leq p + q$,  the
 matrices   associated with   the   $e_i$'s ,  i.e. such that $ (e_i \cdot s)^\a =  \Dirac i \b \a  s^\b$ for any  $s = (s^\a) \in S =\bK^N$. \par
 Recall that there is a natural  {\it vector  space} isomorphism $\varphi: \cC\ell(V) \longrightarrow \L^* V$, which can be used to define   the following
  ``Clifford product''  
$$ \cdot : \L^r V \times S\longrightarrow  S\ ,\qquad  B \cdot s \= \varphi^{-1}(B) \cdot s\ .$$
If $\displaystyle B =\!\!\!\!\! \sum_{j_1 < \ldots < j_r }  B^{j_1 \ldots j_r}e_{j_1}Ê\wedge \ldots \wedge e_{j_r} $ and $s = (s^\a)$,   the components  of  $B\cdot s $  are 
$$ (B \cdot s)^\a \= \sum_{j_1 < \ldots < j_r}B^{j_1 \ldots j_r}
\left(\G_{j_1}Ê\cdot  \ldots \cdot  \G_{j_r} \right)^\a_\b  s^\b\ .$$
Clifford products   between    fields in    $\cD^\perp$ and  fields in  $\cD$ are defined as follows.\par
\begin{definition} A  tensor  field $w \in  \locsect{\L^r \cT M^\LL}$    is called {\it $\cD$-orthogonal} if it takes values  in  the sheaf  generated by  wedge products of vector fields  in $\cD^\perp$.  An $r$-form  
$\o \in  \locsect{\L^r \cT^* M^\LL}$     is called {\it $\cD$-orthogonal} if $\o^\sharp $ is $\cD$-orthogonal. 
The sheaves of $\cD$-orthogonal $r$-forms  and skew-symmetric tensor fields of type $(r,0)$ will be denoted by 
$\Lambda^r \cD^{\perp *}$ and $\Lambda^r \cD^\perp$, respectively.
\end{definition}
 \par
Notice that,  when  $(E_A) = (E_i, E_\a)$ is a frame field of \S \ref{notation},  a  tensor field $w$ is  in $\locsect{\L^r \cD^\perp}$  if and only if it is  the form 
$ w =  \sum_{j_1 < \ldots < j_r}w^{j_1 \ldots j_r}E_{j_1} \wedge \ldots \wedge E_{j_r}$.
Given  $w \in \locsect{\L^r \cD^\perp}$, $s = s^\a E_\a \in \locsect{\cD}$,   we call  {\it Clifford product between  $w$   and  $s$}  the  vector field in $\cD$   
$$w \cdot s \= \left(\sum_{j_1 < \ldots < j_r} w^{j_1 \ldots j_r} \left(\G_{j_1} \cdot \ldots \cdot \G_{j_r}\right)_\b^\a s^\b\right) E_\a\ .$$
By $\SO(V)$-equivariance, $w \cdot s$   does not depend on   the frame field $(E_A)$. \par 
\medskip
\subsection{Orientations, $\cD^\perp$-curvatures and  Rarita-Schwinger 1-form} \hfill\par
\begin{definition} Let   $\cG = ((M^\LL, M_o^\LL, \cD),  (g, \n) )$ be a supergravity of type $\gg$.  A  {\it $\cD^\perp$-volume form}  is an 
even  $n$-form  $\o \in \sect{\L^n \cD^{\perp*}}$,  satisfying  the following condition: for any system of coordinates $\xi = (x^i): \cU_o \subset M_o \longrightarrow \bR^n$ and for any $x \in \cU_o$, the element in $\LL = \L^* W$
$$\l(x) = \o|_{M_o}\left(\left.\frac{\partial}{\partial x^1}\right|_{x}, \ldots, \left.\frac{\partial}{\partial x^n}\right|_{x}\right)$$
is  invertible, i.e. its component $\l^\bR(x)$  in  $\bR \subset \L^*W$ is different from  $0$. 
 If  there is  a  $\cD^\perp$-volume form, we say  that  {\it $\cG$ is orientable}. We call it {\it oriented} when it is  endowed with  a fixed   $\cD^\perp$-volume form $\o$, determined up multiplication by  an invertible $\LL$-superfunction $\l$. 
\end{definition}
Notice that  {\it $\cG$ is orientable if and only  if $M_o$ is orientable}.  \par
\medskip
Given  a   $\cD^\perp$-volume form $\o$, a  frame field $(E_A)$  as  in \S \ref{notation}  is  called  {\it positively oriented} if  there is  a  $\LL$-superfunction $\l$ such  that  $\o Ê= \l \left(E^{1}\wedge \ldots \wedge E^{n}\right) $
with  $\l^\bR|_x > 0$  for any $x \in M_o$. Notice that,  
 for any given $x_o \in M_o$, it always possible  to  determine  a  positively oriented frame field $(E_A)$  on a  neighborhood  $\cU^\LL \subset M^\LL$ of  $x_o$. In the following, when  $M^\LL$ is oriented, the frame fields  are  tacitly assumed to be positively oriented.   \par
\medskip
 Consider now an oriented supergravity  $\cG = ((M^\LL, M_o^\LL, \cD),  (g, \n) )$. We call {\it Hodge star operator}   the linear operator $\ast:\locsect{\L^r \cD^{\perp *}} \longrightarrow \locsect{\L^{n-r}
 \cD^{\perp *}}$ defined as follows. Given a positively oriented frame field $(E_A)$ and a $\cD$-orthogonal $r$-form $w$,   we know that   $w^\sharp$ is   (locally) of the form 
$ w^\sharp =  \sum_{m_1 < \ldots < m_r}w^{m_1 \ldots m_r}E_{m_1} \wedge \ldots \wedge E_{m_r}$.
We define 
$$\ast w = \sum_{m_1 < \ldots < m_r; j_1 < \ldots < j_{n-r}}
\e_{m_1 \ldots m_{r} j_1 \ldots j_{n-r}}  w^{m_1 \ldots m_r} \ E^{j_1} \wedge  \ldots \wedge E^{j_{n-r}}\ .$$
One can check that $\ast w$ is independent of the choice of  $(E_A)$: it depends only on 
$w$, $g$ and the orientation. \par
\medskip
 Let $z$, $z' \in \locsect{\L^r \cD^{\perp*}}$, with components in a frame field
$$  z =  \sum_{m_1 < \ldots < m_r}z_{m_1 \ldots m_r}E^{m_1} \wedge \ldots \wedge E^{m_r} \ ,\qquad z'{}^\sharp = \sum_{j_1 < \ldots < j_r}z'{}^{j_1 \ldots j_r}E_{j_1} \wedge \ldots \wedge E_{j_r}\ .$$
We call {\it inner product between $z$ and $z'$} the $\LL$-superfunction
$g(z, z') \=   z_{m_1 \ldots m_r} z'{}^{m_1 \ldots m_r}$.
Notice that  $g(z, z')$ is independent of the choice of the frame field.
By  little abuse of notation, we   denote  $\| z\|^2_g = g(z, z)$, even if
 ``$\|z\|_g$''  does not exist. \par
\medskip 
In analogy with  \cite{SaS},  we denote by   $g^{\cD^\perp}$,  $\Ricperp$  and $s^{\cD^\perp}$  the even  tensor fields  of type $(0,2)$ and even $\LL$ superfunction 
$$g^{\cD^\perp}(X, Y) = g(\pi^{\cD^{\perp}}(X), \pi^{\cD^\perp}(Y))\ ,$$
$$\Ricperp(X, Y) = \sum_{i = 1}^{n} \e_i g(R_{\pi^{\cD^{\perp}(X)} E_i} \pi^{\cD^{\perp}(Y)},  E_i)\ ,$$
$$ s^{\cD^\perp}=\sum_{j = 1}^n\epsilon_j\Ricperp (E_j,E_j)\ , $$
where $(E_A) = (E_i, E_\a)$ is a frame field as  in \S \ref{notation} and  $\e_i = g(E_i, E_i) = \pm 1$.\par 
Finally,  we call  {\it Rarita-Schwinger 1-form}  the even tensor field $\cR \in \sect{ \cD \otimes \cT^*M^\LL }$ defined by
$$ \cR(X) \=  \sum_{i < j }  \e_i \e_j (\pi^{\cD^\perp}(X) \wedge E_i \wedge E_j) \cdot \left\{ (\pi^\cD \circ T)\left(E_i, E_j\right)\right\}$$
where $ (E_i, E_\a)$ is a frame field as  in \S \ref{notation} and  ``$\cdot$''  denotes a  Clifford product.
\par
\smallskip
 The tensor fields on $M_o$ given by the restrictions  $ \Ricperp|_{ \vee^2 TM^\LL_o}$ and $ \cR|_{T M_o^\LL}$ can be written in terms of   graviton, gravitino,  Ricci tensor of the metric connection $D$  and   covariant derivatives $\bD_{\frac{\partial}{\partial x^i}} \left(\vartheta(\frac{\partial}{\partial x^j})\right)$. To check this,  we refer to \cite{SaS}  since the required
expressions  are formally identical  to those  that one can  derive  on a non-super space-time. \par 
\medskip
\subsection{Supergravity in 11 dimensions}\hfill\par
\label{section52}
According to the remarks  in \S \ref{generalcovariance}, 
{\it all results,  established  in \cite{SaS} for  non-super space-times,  can be re-formulated   in the context of \underline{super} space-times}, provided  that  appropriate adjustments in signs  are  taken into account.  \par
\smallskip
So, as  in \S  4.1 of  \cite{SaS}, if $(g, \n)$ is a gravity field on a super-spacetime $(M^{\LL}, M_o^\LL, \cD)$, the torsion $T$ of $\n$  decomposes  into a sum of the form 
$$ \label{decoT}  T = T^{\cD^\perp} + T^{\cD}_x + \cC^{\cD, \cD^\perp; \cD} + \cC^{\cD, \cD^\perp; \cD^\perp}
 +  \cH^{\Lambda^2\cD^\perp; \cD} +  \cH^{\Lambda^2\cD; \cD^\perp}\ , $$
 with
$$T^{\cD^\perp}   \in \sect{ \cD^\perp \otimes \L^2 \cD^{\perp*}}\ ,\quad T^{\cD}   \in \sect{\cD \otimes \L^2 \cD^*}\ ,$$
$$\cC^{\cD, \cD^\perp; \cD}   \in \sect{\cD \otimes \cD^*\otimes \cD^{\perp*} } \ ,\qquad
\cC^{\cD, \cD^\perp; \cD^\perp}  \in \sect{ \cD^\perp \otimes \cD^* \otimes \cD^{\perp*}}
\ , $$
$$\cH^{\Lambda^2\cD^\perp; \cD} \in \sect{\cD \otimes \L^2\cD^{\perp*}}\ ,\qquad \cH^{\Lambda^2\cD; \cD^\perp} \in\sect{\cD^\perp \otimes \L^2\cD^*}\ ,$$
where  $\cD^{\perp *}, \cD^*\subset \cT^*M^\LL$ are  the sheaves of 1-forms which vanish identically on sections of $\cD$ and $\cD^\perp$, respectively.
Moreover,  as in  \cite{SaS}, 
\begin{itemize}
\item[--]  $ \cH^{\Lambda^2\cD; \cD^\perp} = - \cL$, {\it where $\cL$ is the  Levi tensor $\cL$ of $\cD$, given  by $\cD^\perp$}; 
\item[--] {\it for any tensor field  $g$ on $(M^\LL, M_o^\LL, \cD)$, satisfying Definition \ref{supergravityfields2} (i),  there exists an essentially unique connection $\n$ such that }
\beq \label{defLC} 
T^{\cD^\perp}=0\ \ \text{and}\ \ \ \cC^{\cD, \cD^\perp; \cD^\perp}\in\sect{ \Sym(\cD^\perp)\otimes \cD^*}\ , 
\eeq
{\it where  $\Sym(\cD^\perp)\otimes \cD^*$ is the  sheaf  of  the  sections $C \in \cD^\perp \otimes  \cD^{\perp^*} \otimes \cD^* $ satisfying
$g(C(s, V), V') =  (-1)^{|s| |V| } g(V, C(s, V'))$
for  any homogeneous  $V, V'\in \locsect{\cD^\perp}$, $s \in \locsect{\cD}$} (for detailed statement, see \cite{SaS},  Thm. 4.1).
\end{itemize}
Supergravities  satisfying \eqref{defLC} are  called {\it Levi-Civita}, while we 
call {\it strong Levi-Civita} the supergravities satisfying the following stronger constraints:
\begin{itemize}
\item[1)] $T^{\cD^\perp} = 0 = \cC^{\cD, \cD^\perp; \cD^\perp}$ (i. e.  {\it strict Levi-Civita} according to \cite{SaS}),  
\item[2)]  $T^\cD \equiv 0$.  
\end{itemize}
 We point out that   (1) and (2)  are   the   constraints,  which  appear in superspace formulation of   simple $4D$-supergravity (see \cite{WZ, WB, vN, SaS})  and, as we will shortly see,   play a crucial role in the theory of supergravities in 11 dimensions. \par
\bigskip
Let us now show how the theory of  $11D$-supergravity by  Cremmer, Julia and Scherk (\cite{CJS}) and its (on-shell) superspace formulation (\cite{CF, BH} (see also \cite{CDF})  can be presented in terms  of supergravities of type $\gg$.  \par
Let $V = \bR^{10,1}$ and $\gg = \so(V) + V + S$ the super-Poincar\`e algebra,  determined by the admissible bilinear form  $\b(s,s') =  \Im(is^T \G_0 s')$  on the irreducible module  $S = \bC^{32}$ of Dirac spinors, with Dirac matrices $\G_0$ antisymmetric and $\G_i$, $i \neq 0$, symmetric.  
\par
\begin{definition} We call  {\it CJS-supergravity} a triple $\cG_{CJS} = ((M^\LL, M_o^\LL, \cD)$, $(g , \n)$,  $\cF)$, formed by 
\begin{itemize}
\item[--] an oriented super space-time $(M^\LL, M_o^\LL, \cD)$ of type $\gg$; 
\item[--] a gravity field $ (g , \n)$  on  $M^\LL$; 
\item[--] an even  $\cD$-orthogonal, 4-form $\cF$ on $M^\LL$, 
\end{itemize}
and subjected to the following constraints
  \begin{itemize}
\item[1)]  $\n$  is strong  Levi-Civita (i.e. $T^{\cD^\perp} = 0 = \cC^{\cD, \cD^\perp; \cD^\perp} = T^\cD $);  
\item[2)]   for any even $X\in \locsect{\cD^\perp}$ and odd $s \in \locsect{\cD}$
$$ \cC^{\cD,\cD^\perp;\cD}(s,X) =\frac{1}{144}(X \wedge \cF^\sharp  - 8 (\imath_X \cF)^\sharp ) \cdot s\ ,$$ 
where ``$\cdot$'' denotes Clifford product.
\end{itemize}
If $\cG_{CJS}$ is a  CJS-supergravity, its {\it super-flux} is the skew-symmetric tensor field $F \= \cF + \cZ$ with $\cZ$ defined by 
$$ \cZ(X_1, X_2, X_3, X_4) \=\phantom{aaaaaaaaaaaaaaaaaaaaaaaaaaaaaaaaaaaaaaaaaaaa}$$
$$\phantom{a} =  \frac{1}{8} \sum_{\s \in P_4} \varepsilon(\s, X)g\left( i \pi^{\cD}(X_{\s(1)}), \left(\pi^{\cD^\perp}(X_{\s(2)})\wedge\pi^{\cD^\perp}(X_{\s(3)})\right)\cdot \pi^{\cD}(X_{\s(4)})\right)$$
where  $ \varepsilon(\s, X)$ is the super-sign  \eqref{signpermutation} and  ``$\cdot$''   denotes Clifford product. 
\end{definition}
Given a CJS-supergravity  $\cG_{CJS}$, we consider as  {\it physical fields of  $\cG_{CJS}$} the graviton, the gravitino, the A-field and the metric and spinor connections, defined as for any other supergravity,   plus the {\it flux}  that is   the tensor field   $\bF \in \sect{\L^4 T^* M_o^\LL}$ defined by 
$$
\label{flux}
\bF(X_1, X_2, X_3, X_4) \= \left.F\left(\wh X_1, \wh X_2, \wh X_3,\wh X_4\right)\right |_{M_o}\ ,$$
where  the $\wh X_i \in \locsect{\cT M^\LL}$  are such  $\wh X_i|_{M_o}Ê= X_i$.
Notice that,  by (1), (2)  and definition of $F$, the A-field and the metric and spinor connections   are  completely determined by  $g$, $\vartheta$ and  $\bF$, so that 
the degrees of freedom of  all physical fields  are given only by  these three fields.   \par
\bigskip
We claim that {\it  Cremmer, Julia and Scherk's  theory  can be considered as the  theory of  CJS-supergravities that  solve the system of equations\/}: 
\begin{itemize}
\item[i)] 
$\left\{\begin{array}{l}
 \left.\left(d\cF +   d\cZ\right)\right|_{TM_o\times \dots \times TM_o}Ê = 0\ , \\
\ \\
\left.\left( (d \ast \cF)^{\cD^\perp}  Ê+   \cF \wedge \cF\right)\right|_{TM_o\times \dots \times TM_o} = 0
\end{array}\right. $ \hfill ({\it  Maxwell equations});
\vskip0.2cm
\item[ii)] $\left. \cR \right|_{T M_o} = 0$\hfill
({\it Rarita-Schwinger equations}); 
\vskip0.2cm
\item[iii)] $
\left.\left(\Ricperp \!\!\!\! - \frac{1}{2} s^{\cD^\perp}g^{\cD^\perp}\!\!\!\!\! -  \frac{1}{24}  \left(\|\cF\|_g^2 \ g^{\cD^\perp}\!\!\! - 8 g (\imath_{(\cdot)}\cF , \imath_{(\cdot)}\cF)\right)\right){}_{\phantom{A_{A_{A_A}}}}\!\!\!\!\!\!\!\!\!\!\!\!\!\!\!\right|_{TM_o \times TM_o}\!\!\!\! = 0$\par
\medskip
\hfill({\it Einstein equations}).
\end{itemize}
(here $(\cdot)^{\cD^\perp} $ is the projection onto the space of $\cD$-orthogonal forms).
In fact,  if $\wh g$, $\vartheta$, $\bF$ are graviton, gravitino and flux of a CJS-supergravity  satisfying  (i) - (iii),  then they    satisfy the Euler-Lagrange equations of Cremmer, Julia and Scherk's Lagrangian for 11D-supergravity: just look  at expressions in coordinates of (i) - (iii) and compare them with the equations in  \cite{CF}. \par
 The converse, i.e.  if   $\wh g$, $\vartheta$, $\bF$  satisfy Cremmer, Julia and Scherk's equations,   then they  are  physical fields of a CJS-supergravity satisfying  (i) -  (iii),  can be checked  as follows. We give here only an informal sketch, planning to give   detailed arguments  somewhere else.   \par
\smallskip
For any  CJS-supergravity $\cG_{CJS} = ((M^\LL, M_o^\LL, \cD), (g , \n),  \cF)$,  one may  consider  the  $\SO(V)$-superbundle $\pi: P^\LL \longrightarrow M^\LL$, generated by  orthonormal frame fields as in \S \ref{notation} (\footnote{For brevity, we omit  the definitions of  ``principal superbundles'' and  related notions, but they can be  guessed by  analogy from corresponding classical definitions.}). The superbundle $P^\LL$ is endowed with the  $V + S$-valued,  soldering  1-form $\theta = e_C \otimes_\LL \theta^C = e_i \otimes_\LL \theta^i + e_\a \otimes_\LL \theta^\a$ and the  $\so(V)$-valued, connection 1-form  $\o^\n =  E^B_A  \otimes_\LL   \o^A_B$, corresponding to the covariant derivation  $\n$. Here, $\theta^C$, $\o^A_B$ are $\LL$-valued 1-forms, $(e_C) = (e_i, e_\a)$ is a basis for $V + S$ as in \S \ref{notation}, and  $(E^B_A)$ is the basis  of $\ggl(V + S)$ with elements defined by $E_A^B \cdot e_C = \d^B_C e_A$;  the 1-forms $\o^A_B$ are such that   $\o^\n =  E^B_A  \otimes_\LL   \o^A_B$ takes values in $\so(V) \subset \ggl(V+S)$.\par
As for classical smooth manifolds, the torsion $\wt T$ and curvature   $\wt R$  of $\o^\n$  are $\SO(V)$-equivariant 2-forms on $P^\LL$  and induce on $M^\LL$ the  torsion $T$  and curvature $R$ of $\n$.  There exists also a (uniquely defined)   $SO(V)$-equivariant    4-form $\wt \cF =  \wt F_{i_1 i_2 i_3 i_4}Ê\theta^{i_1} \wedge \theta^{i_2}  \wedge \theta^{i_3} \wedge \theta^{i_4}$ on $P^\LL$, which induces the 4-form  $\cF$  on $M^\LL$. One can check that if   constraints (1),  (2)   and equation $d \cF + d \cZ = 0$ are satisfied, then 
 $\wt T$, $\wt R$ and $\frac{1}{3}\wt \cF$ are the curvatures  of a  Free Differential Algebra $\cA$ satisfying    the constraints given in  \cite{CDF} (III.8.34) - (III.8.37)
 (for  definition of Free Differential Algebras and  applications  to  supergravity, see   \cite{CDF, DF, Fre}).\par
It follows that   $\wt T$, $\wt R$ and $\frac{1}{3}\wt \cF$  satisfy a system of equations, given by   the generalized Bianchi identities of $\cA$ and the integrability conditions, which are   consequences of constraints and Bianchi identities (see \cite{CDF},  p. 908--910).  Neglecting  the  usual Bianchi identities of  $\wt T$  and $\wt R$ (which are automatically satisfied because we are assuming that $\o^\n$  is a connection form) and the relations   identically satisfied because of the constraints,  the system  is equivalent to the following set of equations  on  tensor fields on $M^\LL$:
\begin{itemize}
\item[i')] 
$\left\{\begin{array}{l}
d F = d\cF +   d\cZ  = 0, \\
\ \\
 (d \ast \cF)^{\cD^\perp} Ê+   \cF \wedge \cF\ = 0;
\end{array}\right. $
\vskip0.2cm
\item[ii')] $\cR  = 0$; 
\vskip0.2cm
\item[iii')] $
\Ricperp \!\!\!\! - \frac{1}{2} s^{\cD^\perp}g^{\cD^\perp}\!\!\!\!\! -  \frac{1}{24}  \left(\|\cF\|_g^2 \ g^{\cD^\perp}\!\! \!- 8 g (\imath_{(\cdot)}\cF , \imath_{(\cdot)}\cF)\right)= 0$.
\end{itemize}
Now, we recall that  \cite{CDF} (III.8.34) - (III.8.37)  are  ``rheonomic constraints'' and  that   any  collection of differential forms on  $P^\LL_{\dl M_o}$,  satisfying restrictions of  the above generalized Bianchi identities and   integrability conditions,   admits   a unique  extension to a Free Differential Algebra $\cA$ on  $P^\LL$ satisfying  those constraints (see \cite{CDF}, Ch. III. 3). 
On the other hand,  (i) - (iii) are restrictions to vector fields in $TM_o$ of the equations (i') - (iii') and hence they correspond to restrictions to vector  fields in $T(P^\LL_{\dl M_o})$ of the Bianchi identities and   integrability conditions of $\cA$.   We infer that  the triples $(\wh g, \vartheta, \bF)$ satisfying Cremmer, Julia and Scherk's equations (which, we recall,  are the equation one gets when writes   (i) - (iii) in terms of  $(\wh g, \vartheta, \bF)$)   are in one-to-one correspondence with the   Free Differential Algebras $\cA$ on $P^\LL$, satisfying  the  quoted  rheonomic constraints. Since  such Free Differential Algebras   correspond uniquely to CJS-supergravities satisfying (i') - (iii'), the claim follows. \par
\smallskip
It is important to observe that such arguments show also that a CJS-supergravity satisfies (i) - (iii)  if and only if it satisfies (i') - (iii'). {\it Since  (1), (2) and  (i') - (iii')  are  of tensorial type,    they are  manifestly covariant and  hence they satisfy the Generalized Principle of General Covariance}. \par
\medskip
Finally, as pointed out in \cite{CDF},  we remark that all equations (i') - (iii'), except $d F = 0$,   are integrability conditions that are  automatically satisfied  by any CJS-supergravity  solving  $d F = 0$.  This intriguing fact was first proved   by Brink and Howe and Cremmer and Ferrara  in \cite{BH, CF}. \par
\bigskip
\appendix
\section{Basics of supergeometry}
\label{appendix1}
\setcounter{equation}{0}
\subsection{A digest of supermanifolds} \label{digest} 
\subsubsection{First definitions} \label{firstdef} A  {\it smooth supermanifold  of dimension $(n|m)$}  is  a  pair $M = (M_o, \cA_M)$,  formed by an   $n$-dimensional smooth manifold $M_o$ (called {\it body}) and  a  sheaf of $\bZ_2$-graded 
algebras  $\pi: \cA_M= \cA_{M0} + \cA_{M1} \longrightarrow M_o$ (called {\it sheaf of superfunctions}), such that  
\begin{itemize}
\item[--]{\it  there exists an open covering $\{\cU_{oJ}\}$ of $M_o$ with the property that for any restriction $\cA_M{}_{\dl \cU_{oJ}}$ there exists a trivial vector bundle $\pi: S_J \longrightarrow \cU_{oJ}$ of rank $m$  such that $\cA_M{}_{\dl \cU_{oJ}} \simeq \sheaf( \L S^*_J)$.}
\item[--] {\it the sheaf $\pi: \cA_M/(\cA_{M1} + \cA_{M1}^2)\longrightarrow M_o$ is  isomorphic with the sheaf $\gF_{M_o}$ of germs of smooth real functions on $M_o$}.
\end{itemize}
Any  pair $\cU = (\cU_o, \cA_M{}_{\dl \cU_o})$, with an isomorphism $\cA_M{}_{\dl \cU_o}  \simeq \sheaf( \L S^*)$ for a trivial vector bundle $\pi:S \to \cU_o$, is called {\it decomposable neighborhood of $M$}. \par
\medskip
Let $\cU = (\cU_{o}, \cA_M{}_{\dl \cU_{o}})$ be a decomposable neighborhood and assume that 
there exists coordinates  $\xi = (x^i): \cU_o \longrightarrow \cU'_o = \xi(\cU) \subset \bR^n$, which we  use to make the identifications  $\cU_o \simeq \cU'_o \subset \bR^n$ and $S \simeq \bR^m \times \cU'_o$. The corresponding   isomorphism $\wh \xi: \sheaf{(\L \bR^m{}^* \times \cU'_o)} \longrightarrow \cA_M{}_{\dl \cU_o}$ will be called {\it system of supercoordinates 
on $\cU$ associated with $\xi = (x^i)$.}\par
Denoting by $(e^\a)$  the standard basis of $\bR^m{}^*$ and by $\vartheta^\a:  \cU'_o \longrightarrow (\bR^m)^* \times \cU'_o$ the constant sections $\vartheta^\a(x) \equiv e^\a$, the superfunctions $\gf \in \sect{ \cA_M{}_{\dl\cU_o}}$  can be identified with the   sections  of $\sheaf{(\L \bR^m{}^* \times \cU'_o)}$ 
 \beq \label{superf}\gf = \sum_{ \begin{smallmatrix}
\a_j = 0,1 \end{smallmatrix}} \!\!\!\! \gf_{\a_1 \dots \a_m} (x^1, \dots , x^n) (\vartheta^1)^{\a_1} \wedge\dots \wedge  (\vartheta^m)^{\a_m}  \ ,\eeq
where  $(\vartheta^\b)^0 = 1$ and $1 \wedge \vartheta^\b = \vartheta^\b$. 
 For simplicity,  we set $(\vartheta^\b)^\a = 0$
for any  $\a \neq 0,1$, so that  \eqref{superf}   makes sense even if we sum over  $\a_j \in \bN$.\par
  \medskip
 The homogeneous superfunctions $x^i$ and  $ \vartheta^\a$  are called 
{\it even\/} and {\it odd  coordinates}, respectively. One can check that
even (resp. odd) superfunctions, i.e.  superfunctions  of parity $0$ (resp. $1$), are of the form 
$$ \label{evenfunction} \gf = \sum_{ \begin{smallmatrix}
\sum_j \a_j = 0\!\!\!\!\mod{2} \end{smallmatrix}} \!\!\!\! \gf_{\a_1 \dots \a_m} (x^1, \dots , x^n) (\vartheta^1)^{\a_1} \wedge\dots \wedge  (\vartheta^m)^{\a_m}  \ ,$$
$$\left(\ \ \gf = \sum_{ \begin{smallmatrix}
\sum_j \a_j = 1\!\!\!\!\!\mod{2}  \end{smallmatrix}} \!\!\!\! \gf_{\a_1 \dots \a_m} (x^1, \dots , x^n) (\vartheta^1)^{\a_1} \wedge\dots \wedge  (\vartheta^m)^{\a_m}\ \right).\qquad\quad\,\,$$
\par
\medskip
A  {\it morphism\/}  between supermanifolds $M = (M_o, \cA_M)$ and $N = (N_o, \cA_N)$ is a pair 
 $(f, \wh f)$ formed by a smooth map $f: M_o \longrightarrow N_o$ and a morphism $\wh f: \cA_N \to f_*(\cA_M)$
of sheaves of $\bZ_2$-graded algebras over $N_o$.  
\par
\medskip
For any supermanifold $M = (M_o, \cA_M)$, one can check that the subsheaf  $\gJ_M \subset \cA_M$ of  nilpotent superfunctions, which is generated by germs of the form
$ \gf =   \sum_{ \a_1 +  \dots + \a_m \geq 1} \gf_{\a_1 \dots \a_m} (x) (\vartheta^1)^{\a_1} \wedge\dots \wedge  (\vartheta^m)^{\a_m}$, coincides with the sheaf $(\cA_{M1} + \cA_{M1}^2)$ and hence that   $\cA_M/\gJ_M$ is identifiable with $\gF_{M_o}$. 
\par
The natural projection 
$\epsilon: \cA_M  \longrightarrow  \cA_M/\gJ_M\simeq \gF_{M_o}$ is called {\it evaluation map}. For a superfunction of the form
\eqref{superf}, 
$$ \label{eval}Ê\e(\gf) = \gf_{0\dots 0}(x^1, \dots, x^n)\ .$$
This fact    is often described saying that  ``$\e(\gf)$ is the function given  by evaluating   \eqref{superf}  at  $\cU_o = \{\ \vartheta^\a = 0\ \}$''. According to this, for any $\gf \in \locsect{\cA_M}$ and $x \in M_o$,  we adopt the notation  
$$\gf|_{M_o} = \e(\gf)\qquad\text{and}\qquad  \gf|_x \ \text{or} \ \gf(x) =    \e(\gf)(x)\ .$$
The morphism $\imath_{M_o} = (Id_{M_o}, (\cdot)|_{M_o}): (M_o, \gF_{M_o}) \longrightarrow M = (M_o, \cA_M)$ is called {\it natural embedding of $M_o$ into $M$}.\par
\medskip
A supermanifold of dimension $(0|0)$ and connected body is called {\it superpoint}. It is unique up to isomorphism and is denoted by  $\bR^{0|0}$.  Any $x \in M_o$   is naturally identified with the superpoint $(\{x\}, \bR) \simeq \bR^{0|0}$. The {\it natural embedding of $x$ in $M$} is the morphism  $\imath_{x} = (Id_{x}, (\cdot)|_{x}): (\{x\}, \bR) \longrightarrow M$.\par
  \medskip
  \subsubsection{Cartesian products of supermanifolds}\label{cartsup}
  Let $M_i = (M_{io}, \cA_{M_i})$, $i = 1,2$,  be two  supermanifolds and  $\pi_i: M_{1o} \times M_{2o} \longrightarrow M_{io}$ the natural projection of  $M_{1o} \times M_{2o}$   onto the $i$-th factor.  The {\it Cartesian product of $M_1$ and $M_2$} is the supermanifold given by the pair
$$ M_1 \times M_2 \= (M_{o1} \times M_{o2}, \cA_{M_1 \times M_2})\ ,$$
 where 
$\pi: \cA_{M_1\times M_2} \longrightarrow M_{o1} \times M_{o2}$ is a sheaf, which is canonically determined by $\cA_{M_1}$, $\cA_{M_2}$ and  includes 
$ \pi^*_1(\cA_{M_1}) \otimes  \pi^*_2(\cA_{M_2})$ as a dense subsheaf (for a detailed definition of $\cA_{M_1\times M_2}$, see  \cite{Ko}, p. 215). \par
For any $x \in M_{o1}$, the {\it evaluation of superfunctions of $M_1\times M_2$ at $x$} is the sheaf morphism $\e_x: \cA_{M_1 \times M_2} \longrightarrow \cA_{M_2}$ defined by 
$$\e_x(\ga \otimes \gb) =  \ga|_x\  \gb\ \qquad  \text{for any} \ \ga \in \locsect{\cA_{M_1}}\ ,\ \ \gb \in \locsect{\cA_{M_2}}\ .$$
\par
\medskip
 \subsubsection{Tensor fields} \label{supervect} The {\it supervector fields} of a supermanifold $M$ $= (M_o$, $\cA_M)$ (shortly called {\it vector fields}) are the derivations of $\sect{\cA_M}$.  In supercoordinates, they correspond to derivations of  $\sect{\cA_M{}_{\dl \cU_o} }$ $=$ $\sect{\sheaf{(\L \bR^m{}^* \times \cU_o')}}$ of the form
 $$X = X^j \frac{\partial}{\partial x^j} + X^\a \frac{\partial}{\partial \vartheta^\a}\ ,\qquad X^j\ ,\  X^\a \in \sect{\cA_{M \dl\cU_o}}\ ,$$
 where $\frac{\partial}{\partial x^j}$, $\frac{\partial}{\partial \vartheta^\a}$ are  such that
$\frac{\partial}{\partial x^j}x^k = \d^k_j$,   $\frac{\partial}{\partial \vartheta^\a}\vartheta^\b = \d^\b_\a$,  $\frac{\partial}{\partial x^j}\vartheta^\a  =  \frac{\partial}{\partial \vartheta^\a} x^k = 0$. \par
\medskip
The sheaf $\pi: \cT M \longrightarrow M_o$ of germs of  vector fields  is called {\it tangent  sheaf}. It has a natural structure of sheaf of $\bZ_2$-graded $\cA_M$-modules. The vector  fields  $\frac{\partial}{\partial x^j}$, $\frac{\partial}{\partial \vartheta^\a}$ have parity $0$ and $1$, respectively. \par
 \medskip
  The {\it Lie bracket} of  homogeneous  $X$, $Y \in \sect{\cT M}$ is  defined by
 \beq 
 \label{brackets}[X, Y]\cdot f \= X\cdot (Y\cdot f) - (-1)^{|X| |Y|} Y\cdot (X\cdot f)\ ,\quad \text{for any} \ f \in \locsect{\cA_M}\ . \eeq
 This operation is extended  $\bR$-bilinearly on arbitrary  pairs $X, Y \in  \sect{\cT M}$.\par
 \medskip
 For any $x \in M_o$, the {\it tangent space of $M$ at $x$} is the $\bZ_2$-graded vector space $T_{x} M = (T_{x} M)_0 + (T_{x} M)_1$,  with   $(T_{x} M)_\a$
 defined by 
 $$ (T_{x} M)_\a \= \{\ v:\cA_M{}_{\|{x}}\rightarrow\bR\ :\ \ \ v(\gf \gg) = v(\gf) \gg|_x + (-1)^{\a |\gf|} \gf|_x v(\gg) \quad \text{and}\phantom{aaaaaa}$$
 $$\phantom{aaaaaaaaaaaaaaaaaaaaaaaaaaaaaaaa} v(\gf) = 0\  \text{for any}\ \gf\in\cA_{M [\a +1]_{\!\!\!\!\!\!\mod \!2}} \  \}\ ,$$
 \par
 \noindent
 We denote the bundle $\pi: \bigcup_{x \in M_o} T_{x} M \longrightarrow M_o$ by   ``$T M|_{M_o}$'' 
(\footnote{In the literature,  $TM|_{M_o}$ is  usually 
  denoted by ``$TM$''.  We  decided to use such new notation, because  $TM|_{M_o}$  is similar more  to the  restriction of a tangent bundle  to a submanifold than to the  tangent bundle of a manifold.}).  \par
 \medskip
%
%
 It is known (see e.g. \cite{Ko}, \S 2.12) that   $\pi: \sheaf{(T M|_{M_o} )} \longrightarrow M_o$ is isomorphic to  the sheaf determined by   the pre-sheaf  
 $$\{\ \cU_o \longrightarrow \operatorname{Der}(\sect{\cA_{M\dl{\cU_o}}}, \cC^{\infty}_{M_o}(\cU_o) )\  \}\ .$$ 
On the base of  such isomorphism, the evaluation map $\epsilon: \cA_M \to \gF_{M_o}$  determines  a surjective map $\pi^\e: \cT M \longrightarrow \sheaf{(T M|_{M_o})}$  defined by   
$$ \label{pie} \pi^\epsilon(X)\cdot \gf \= \epsilon(X\cdot \gf)\ .$$
If we consider  supercoordinates $(x^i, \vartheta^\a)$ 
on a decomposable neighborhood $(\cU_o,\cA_{M\dl\cU_o})$ 
and set  
$\left.\frac{\partial}{\partial x^i}\right|_{M_o} \= \pi^\e\left(\frac{\partial}{\partial x^i}\right)$, $ \left.\frac{\partial}{\partial \vartheta^\a}\right|_{M_o} \= \pi^\e\left(\frac{\partial}{\partial \vartheta^\a}\right)$, we have that
$$\pi^\e\left( X^j \frac{\partial}{\partial x^j} + X^\a \frac{\partial}{\partial \vartheta^\a}\right) = X^j|_{\cU_o} \left.\frac{\partial}{\partial x^j}\right|_{M_o} + X^\a|_{\cU_o} \left.\frac{\partial}{\partial \vartheta^\a}\right|_{M_o}\ .$$
For any vector field $X \in\G( \cT M)$, we use the notation
$$ X|_{M_o} \= \pi^\e(X)\ ,\ \ \  X|_x \=\pi^\e(X)|_x  \in T_xM\ ,\qquad \ x \in M_o\ ,$$
and we say that {\it $X$ is tangent to $M_o$} if 
for any $x \in M_o$ and any system of super-coordinates, one has that $X|_x$ is of the form 
$X|_x = X^i\left. \frac{\partial}{\partial x^i}\right|_x$ for $X^i\in\bR$.
\par
\medskip
For any morphism $\varphi = (f, \wh f): M = (M_o, \cA_M) \longrightarrow N = (N, \cA_N)$, we denote by $\varphi_*:  f_* \cT M \longrightarrow \Der_\varphi(\cA_N,f_*\cA_M)$ the sheaf morphism 
defined by 
$$\varphi_*(X) \cdot \gf \= X \cdot (\wh f(\gf))\ \text{for any} \ \gf \in \locsect{\cA_N}\ ,\ X \in \locsect{f_* \cT M}\ .$$
\par
\medskip
We conclude with the  definition of   tensor fields.  The {\it cotangent sheaf of $M$}  is the sheaf 
$$\pi: \cT^* M \= \Hom_{\cA_M}(\cT M, \cA_M)\longrightarrow M_o\ .$$
 A section $\o$ of $\cT^* M$ is called {\it 1-form}. It is {\it  homogeneous of parity $|\o|\in\bZ_2$} if 
$$\o(\cT M_i)\subseteq \cA_M{}_{i+|\o|}\qquad \text{and}\qquad \o(\gf X) = (-1)^{|\o| |\gf|} \gf \o(X)$$
for any homogeneous $\gf \in \locsect{\cA_M}$ and $X \in \locsect{\cT M}$. 
\par
\smallskip
The {\it full tensor sheaf of $M$} is the sheaf 
$$\pi: \otimes_{\cA_M}\!\!\!\! < \cT M, \cT^* M> \longrightarrow M_o\ ,$$
  generated by the tensor products ($\bZ_2$-graded over $\cA_M$) of $\cT M$ and $\cT^* M$.  A local section $\a$  of $\otimes_{\cA_M}\!\!\!\! < \cT M, \cT^* M> $ is called {\it tensor field}. It  is  {\it of type $(p,q)$} if it is  sum of tensor products of $p$ vector fields and $q$ 1-forms. It is called 
 {\it homogeneous of parity $|\a| \in \bZ_2$\/}  if it is  sum of  tensor products of homogeneous vector fields and homogeneous $1$-forms, whose sum (in $\bZ_2$) of parities is equal to $|\a|$. 
\par
\medskip
\subsubsection{Skew-symmetric tensor fields, exterior differentials and interior multiplications} 
\label{p-forms}
  A tensor field $\o$ of type $(0,q)$ of $M$ is called {\it  symmetric} {\it ({\rm resp.} skew-symmetric)\/} {\it in graded sense} if for any $q$-tuple $X_1, \dots, X_q$ of homogeneous vector fields and any $1 \leq i \leq q-1$
$$\o(X_1, \dots, X_{i}, X_{i+1}. \dots, X_q) =  (-1)^{|X_i| |X_{i+1}|} \o(X_1, \dots, X_{i+1}, X_i, \dots, X_q)$$
$$\left( \text{resp.} = - (-1)^{|X_i| |X_{i+1}|} \o(X_1, \dots, X_{i+1}, X_i, \dots, X_q) \right)\ .$$
Skew-symmetric $(0,q)$-tensor fields  are  also called {\it $q$-forms}.
Similar definitions are  given for symmetric and skew-symmetric  $(p,0)$-tensor fields  in graded sense.  For brevity,   the words ``in graded sense''  are often  omitted.   \par 
The sheaves over $M_o$, generated by skew-symmetric   tensor fields of type $(p,0)$ and $(0,q)$,  are denoted by 
$\Lambda^p \cT M$ and $\Lambda^q \cT^*M$, respectively, and 
$$\Lambda^* \cT M \= \bigoplus_{0}^\infty \Lambda^p \cT M\ ,\qquad \Lambda^* \cT^* M \= \bigoplus_{0}^\infty \Lambda^p \cT^* M\ .$$
 For an open subset $\cU_o \subset M_o$, the space   $\G\left(\Lambda^* \cT^*M_{|\!|\cU_o}\right)$  is endowed with  a ``wedge  product''
$$\wedge: \G\left(\Lambda^q \cT^*M_{|\!|\cU_o}\right) \times \G\left(\Lambda^{q'} \cT^*M_{|\!|\cU_o}\right)
\longrightarrow \G\left(\Lambda^{q+q'} \cT^*M_{|\!|\cU_o}\right)\ ,$$
which, in the context of  $\bZ_2$-graded multilinear maps on   $\bZ_2$-graded vector spaces,  is the   analogue   of  wedge products  on classical smooth manifolds: differences in the expressions  concerns   only signs, which  have to be consistent with    grades of  arguments and   maps. We refer to \cite{Ko}, p. 244-246, for a detailed  definition of ``$\wedge$''. We point out that such  wedge products  determines  a  natural structure on  $\Lambda^* \cT^* M$ of sheaf of  bi-graded commutative algebras. \par
\smallskip
A corresponding definition determines   a ``wedge product''   between sections in $\Lambda^* \cT M_{|\!|\cU_o}$ and  determines  a natural  structure of sheaf of  bi-graded commutative algebras on $\Lambda^* \cT M$. 
\par
\medskip
For any homogeneous superfunction $\gf \in \locsect{\cA_M}$, the differential $d\gf$ is the 1-form, defined (in analogy with   the classical case)  by 
$$d \gf(X) \= (-1)^{|X||\gf|} X \cdot \gf\qquad \text{for any homogeneous} \ X \in \locsect{\cT M}\ .$$
It can be checked (\cite{Ko}, p.249--250) that, for any open subset $\cU_o \subset M_o$, 
 there exists a unique derivation  $d$ on $\G\left(\Lambda^* \cT^*M_{\dl \cU_o}\right)$ of bidegree $(1,0)$ such that: a) it  coincides with    the differential,  when applied to homogeneous superfunctions; b)  it satisfies   $d^2 =  0$. 
 Such derivation is called {\it exterior differential} and it is the  analogue of the exterior differential of smooth manifolds.  See  \cite{Ko} for its explicit expression  and main  properties. \par
For any open subset $\cU_o \subset M_o$ and $X \in \sect{\cT M_{\dl \cU_o}}$ we denote by 
$\imath_X$  the {\it interior multiplication  by} $X$, i.e. the 
derivation of $\G\left(\Lambda^* \cT^*M_{\dl\cU_o}\right)$ of bidegree $(-1, |X|)$ defined by 
$$  \imath_X \o(Y_1, \dots, Y_p) \= \o(X, Y_1, \dots, Y_p)\ .$$
\par
\medskip
We conclude recalling the definition  of   ``super-signs''  of  permutations of $m$ elements,   frequently  used    
in constructions of skew-symmetric tensors. For any  $\s \in P_m$, we set  
$\D_\s = \{\ (i, j) \ : \ 1 \leq i < j \leq m\ ,\ \s(i) > \s(j)\ \}$ and, if  $X = (X_1, \ldots, X_m)$ is an  $m$-tuple of  homogeneous vector fields $X_i \in \locsect{\cT M}$, we call  {\it super-sign of the pair $(\s, X)$}  the value 
\beq\label{signpermutation} \varepsilon(\s, X) \= (-1)^{\sum_{(i,j) \in \D_\s} (1 + |X_i| |X_j|)}\ .\eeq
When all the $X_i$ are  even, the super-sign coincides to the classical sign $ \varepsilon(\s)$.\par 
\medskip
\subsection{Lie supergroups and Lie  superalgebras}\label{appendixA2}\hfill\par
Given a supermanifold $M = (M_o, \cA_M)$, we   denote by  
$\D_M = (\D_{M_o}, \wh \D_M): M \longrightarrow M \times M$
the {\it diagonal morphism}, determined by  
$$\D_{M_o}(x) = (x, x)\ ,\qquad \wh \D_M\circ \wh \pi_i (\gf) = \gf$$
where $\pi_i: M \times M \to M$, $i = 1,2$,  are the natural projections. \par
\smallskip
A {\it Lie supergroup} is  a supermanifold $G=(G_{o},\mathcal{A}_{G})$, with  body  given by  a Lie group  $G_o$ (whose multiplication map is denoted by $m: G_o \times G_o \longrightarrow G_o$, inversion map by  $n: G_o \longrightarrow G_o$ and  identity by  $e \in G_o$)  and 
 endowed with morphisms
 $\mu=(m,\wh m):G\times G\longrightarrow G$ and $ \nu =(n,\wh n):G\longrightarrow G$,
satisfying the following properties: 
\begin{itemize}
\item[1)] ({\it associativity\/}) as morphisms from $G \times G \times G$ to $G$
$$\mu\circ(Id_{G}\times \mu)=\mu\circ(\mu\times Id_{G})\ ;$$
\item[2)]  ({\it existence of neutral element})  setting ${\bf e} = (\{e\}, \bR)$, 
the following equalities of   morphisms,  from $G \times {\bf e}$ to  $G$ and from ${\bf e} \times G$ to  $G$,   hold:
$$\mu\circ(Id_{G} \times \imath_{{ e}}) =\pi_1\ ,  \qquad \mu\circ(\imath_{{e}}\times  Id_{G} ) =\pi_2\ ;$$ 
\item[3)] ({\it inverse elements\/}) as morphism from $G$ into $G$ 
$$\mu\circ ( Id_{G} \times \nu)  \circ\D_G  = \mu\circ (\nu \times  Id_{G})  \circ\D_G= (e, \{f \longmapsto f(e)\})\ ,$$
where $e: G_o \to G_o$  denotes the constant map with value $e \in G_o$.  
\end{itemize}
\par
A {\it Lie sub-supergroup} of $G = (G_o, \cA_G)$ is a submanifold $H = (H_o, \cA_H)$ of $G$ (i.e. a supermanifold endowed with an embedding $\imath = (\imath_o, \wh \imath): H \longrightarrow G$), whose body is given by a Lie subgroup $\imath_o: H_o \to G_o$ of $G_o$ and such that $\mu' = \mu \circ (\imath \times \imath)$ and $\nu' = \nu \circ \imath$  determine  a structure of Lie supergroup on $H$.\par
\smallskip
For any vector field  $X \in \sect{\cT G}$  of a Lie supergroup of $G = (G_o, \cA_G)$, let us denote by $Id \otimes X$ the corresponding  derivation of $\sect{\cT (G \times G)}$ that acts  only on the second component. The field $X$ is called {\it left-invariant} if it satisfies the condition $(Id \otimes X) \circ \wh m = \wh m \circ X$ (\footnote{This definition  reduces to the usual definition of  ``left-invariant vector fields''  when  $G = G_o$ is a (non-super) Lie group.}).  The  space of left-invariant vector fields, endowed with the brackets  \eqref{brackets}, is a Lie superalgebra, called the {\it Lie superalgebra of $G$}.\par
\begin{definition} A {\it super Harish-Chandra pair} (shortly {\it sHC-pair}) is a pair $(G_o, \gg)$, formed by 
a Lie group $G_o$ and a Lie superalgebra $\gg = \gg_0 + \gg_1$ with $\gg_0=Lie(G_o)$, endowed with a Lie group morphism $\Ad: G_o \longrightarrow \Aut(\gg)$ such that 
\begin{itemize}
\item[--] $\left.\Ad(\cdot)\right|_{\gg_0}\!\!: G_o  \longrightarrow \Aut(\gg_0)$ is  the usual adjoint action of $G_o$; 
\item[--] $\Ad_*\!: \gg_0 \longrightarrow \aut(\gg)$  coincides with  $\ad|_{\gg_0}: \gg_0 \longrightarrow  \aut(\gg)$.
\end{itemize}
\end{definition}
Given a Lie supergroup $G = (G_o, \cA_G)$, one can naturally associate to it the sHC-pair $(G_o, \gg = Lie(G))$ and {\it any sHC-pair corresponds to a unique (up to isomorphism) Lie supergroup}. In particular, given a sHC pair $(G_o, \gg)$, the sheaf $\cA_G$ of the corresponding Lie supergroup $G = (G_o, \cA_G)$ is the  one determined by the pre-sheaf  on $G_o$ 
$$\{\ \cU_o \longrightarrow \Hom(U(\gg), \cC^\infty_{G_o}(\cU_o))^{U(\gg_0)}\ \}\ ,$$
where $U(\gg_0)$ and $U(\gg)$ denote  the universal enveloping algebras of $\gg_0$ and $\gg$, respectively. We refer to \cite{Ks} for the explicit expressions of the product and inverse morphisms $\m$,  $\nu$ of $G = (G_o, \cA_G)$. \par
\bigskip
Let $H=(H_o,\cA_H)$ be a Lie sub-supergroup of   $G=(G_o,\cA_G)$ with  $H_o$ closed in $G_o$. Denote  by $p_o:G_{o}\rightarrow G_{o}/H_{o}$, 
$$
 p = (p_o, \wh p):G\times H\rightarrow G\quad \text{and}\quad \mu_H = (m|_{G_o\times H_o}, \wh m_H): G\times H\rightarrow G
$$
the canonical projections and the restriction of the product rule  $\mu = (m, \wh m): G \times G \to G$ to $G \times H$. The sheaf $\cA_{G/H}$ on $G_o/H_o$, determined by the pre-sheaf 
$$\left\{\ \cU_o\longrightarrow\left\{\gf\in \mathcal{A}_{G\dl \pi_{o}^{-1}(\cU_o)}\ : \ \wh m_H(\gf)=\wh p(\gf)\ \right\}\ \right\}\ ,
$$
 is such that  $G/H=(G_{o}/H_{o},\mathcal{A}_{G/H})$ is a supermanifold, called   {\it homogeneous supermanifold of $G$ modulo $H$\/}. The supergroup  $H$ is called {\it isotropy of $G/H$}. \par

\end{document}